# Fingerprinting the contribution of colored scalars to the $H^+W^-Z(\gamma)$ vertex

**Nabarun Chakrabarty**[1,2,a], **Indrani Chakraborty**[3,b], **Dilip Kumar Ghosh**[4,c]

[1] Centre for High Energy Physics, Indian Institute of Science, C. V. Raman Avenue, Bangalore 560012, India
[2] Physics Division, National Center for Theoretical Sciences, Hsinchu 30013, Taiwan, ROC
[3] Department of Physics, Indian Institute of Technology Kanpur, Kanpur, Uttar Pradesh 208016, India
[4] School of Physical Sciences, Indian Association for the Cultivation of Science, 2A & 2B, Raja S.C. Mullick Road, Jadavpur, Kolkata 700032, India

**Abstract** Color-octet scalars arise in various Grand Unification scenarios and also in other models of new physics. They are also postulated for minimal flavour violation. Purely phenomenological imprints of such scalars are therefore worth looking at. Motivated by this, we perform a complete one-loop calculation of the $H^+ \to W^+Z(\gamma)$ decay in a two Higgs doublet model augmented by a color-octet $SU(2)_L$ scalar doublet. The computation is conveniently segregated into colorless and colored components. The color-octet part of the amplitude, being scaled by the color-factor, provides an overall enhancement to the form factors. Crucial constraints from perturbative unitarity, positivity of the scalar potential, oblique parameters, Higgs signal strengths and direct search of a charged Higgs and color-octet scalars are folded-in into the analysis. Sensitivity of the loop-induced $H^+ \to W^+Z(\gamma)$ vertex to other model parameters is elucidated. Finally, the prospect of observing a loop-induced $H^+ \to W^+Z(\gamma)$ interaction at the future hadronic collisions is also discussed.

**Contents**

## 1 Introduction

The discovery of a Higgs boson at the Large Hadron collider (LHC) [1,2] completes the particle spectrum of the standard model (SM). Further, the interaction strengths of the discovered boson with the SM fermions and gauge bosons are found to be in agreement with the corresponding SM values. However, issues such as a non-zero neutrino mass, the existence of dark matter (DM), the observed imbalance between matter and antimatter in the universe, and, the instability (or metastability) of the electroweak (EW) vacuum in the SM [3–6] hint towards additional dynamics beyond the SM. Interestingly, extending only the Higgs sector of the SM appropriately can suffice to address all the aforementioned issues. Moreover, a scalar of mass 125 GeV with interactions mimicking those of the SM Higgs can be extracted out of such extended Higgs sectors by virtue of additional symmetries or appropriate fine tuning. This therefore motivates putting forth extended Higgs sectors as prototypes of beyond-the-SM (BSM) physics.

[a] e-mail: chakran@iisc.ac.in (corresponding author)
[b] e-mail: indranic@iitk.ac.in
[c] e-mail: tpdkg@iacs.res.in

A two-Higgs doublet model (2HDM) [7,8] is one of such extensions of the SM Higgs sector. It potentially shuts off the flavour changing neutral currents (FCNC), predicts additional CP-violation through the scalar potential that can eventually explain the observed matter-antimatter imbalance, and, poses a solution to the strong CP problem. A 2HDM is in fact the smallest $SU(2)_L$ multiplet to predict a singly charged Higgs $H^+$. An $H^+$ has been searched at the LHC in different channels with its fermionic decay modes being more popular in this context. The $H^+ \to t\bar{b}$ mode is used to look for an heavy $H^+$ whereas the preferred search channel for a light one is the $H^+ \to \bar{\tau}\nu_\tau$. However, probes in such channels are generally swamped by a heavy QCD background. An alternative therefore is to search for its bosonic decays $H^+ \to W^+h, W^+Z, W^+\gamma$. That said, the last two of the aforementioned modes are *prima facie* more intriguing since they arise only at one-loop in multi-Higgs doublet models. The absence of the $H^+W^-Z\gamma$ coupling at the tree level is an artefact of the isospin symmetry of the kinetic terms of the Higgs sector. Since both these characteristics are, in general, broken at one loop level through effects from other sectors that do not respect the custodial invariance, these vertices are induced at the loop level. Momentum dependent interactions appear therein consequently. It therefore gets clear that the strength of the $H^+W^-Z$ interaction captures the custodial symmetry breaking effects in the model embedding it.

An also interesting extension of the SM Higgs sector first proposed in [9] consists of a scalar multiplet transforming as (8, 2, 1/2) under the SM gauge group. The motivation for the same is two-fold. The first is minimal flavour violation (MFV), which is a framework for having flavour-dependent masses without introducing unwanted flavour changing neutral currents (FCNCs). It assumes all breaking of the underlying approximate flavour symmetry of the SM is proportional to the up- or down-quark Yukawa matrices. It has been shown in [9] that the only scalar representations under $SU(3)_c \times SU(2)_L \times U(1)_Y$ complying with MFV are (1,2, $\frac{1}{2}$) and (8,2, $\frac{1}{2}$). Secondly, color-octet scalars can emerge from a plethora of BSM scenarios. Grand unification models discussed in [10–13] contain color-octet scalars. Other examples include topcolour scenarios [14], models with extra dimensions [15,16] and chiral color models. Loop effects of the isodoublet color-octet were looked at in [17–21] More on the TeV-scale phenomenology of color-octet scalars can be found in [22–31]. In fact, an (8,2,1/2) scalar multiplet proved handy in explaining certain anomalous results seen at the Tevatron [32,33] and the Runs I [20,34,35] and II [36,37] of the LHC that had stirred up excitement in those times.

The interest in non-minimal Higgs sectors led to the proposing of a hybrid scenario combining a 2HDM with a color-octet scalar multiplet [38]. Relevant theoretical and experimental constraints were used in [38] to carve out an allowed parameter region. The more stringent requirements of high scale perturbative unitarity and vacuum stability under renormalisation group were imposed in a subsequent study [39]. Since the electroweak interactions of all the scalars in this framework are similar to those of an $SU(2)_L$ doublet, an $H^+W^-Z(\gamma)$ interaction would arise radiatively. Moreover, additional enhancement could be expected due to the color factor. We therefore aim to estimate the strength of the one-loop $H^+W^-Z(\gamma)$ vertex in this model. While a similar calculation was done for the minimal supersymmetric standard model (MSSM) [40,41], a $\mathbb{Z}_2$ symmetric 2HDM [42,43], an aligned 2HDM [44] and a particular version of 3HDM containing two *active* and one *inert* doublet [45], we lay particular emphasis on the contribution coming from the color-octet scalars. The main features of the present study are outlined below.

- To our understanding, this is the first investigation of the impact of colored scalars on the $H^+W^-Z(\gamma)$ vertex. While computing the one-loop amplitude coming from the color-octet, an enhancement by a color-factor is expected. In tandem, also expected is an exclusion limit on the color-octet mass scale from direct searches at the LHC that will tend to dilute this enhancement. We probe the interplay of the two aforementioned effects here.
- We adopt the non-linear gauge to get rid of unphysical vertices involving goldstones. This ends up simplifying the calculation to a good extent. Further, we present certain simplified expressions for the one-loop form factors that make decoupling/non-coupling of the colored scalars from the $H^+W^-Z(\gamma)$ vertex apparent.
- As a phenomenological application, we also discuss how such new contributions change the decay branching fractions of the $H^+ \to W^+Z(\gamma)$ mode, and, consequently, the production cross sections involving these decay processes at the LHC.

The paper is organised as follows. The model is introduced in Sect. 2. A detailed discussion on the chosen non-linear gauge and the analytic calculation of the various form factors is reported in Sect. 3. The constraints applicable to this scenario are elaborated in Sect. 4 and the numerical values of the form factors and branching ratios obtained are reported in Sect. 5. Section 6 outlines the observability of the $H^+W^-Z$ vertex at the hadron colliders. The study is summarised in Sect. 7. Various important formulae are relegated to the Appendix.

## 2 Model description

The model considered replaces the scalar sector of the SM by three $SU(2)_L$ scalar doublets: two color singlets $(\Phi_1, \Phi_2)$ and one color-octet $S$. The most general scalar potential can be written as [38]:

$$
\begin{aligned}
& V(\Phi_1, \Phi_2, S) \\
& = m_{11}^2 \Phi_1^\dagger \Phi_1 + m_{22}^2 \Phi_2^\dagger \Phi_2 - m_{12}^2 \left(\Phi_1^\dagger \Phi_2 + \Phi_2^\dagger \Phi_1\right) \\
& + \frac{\lambda_1}{2} \left(\Phi_1^\dagger \Phi_1\right)^2 + \frac{\lambda_2}{2} \left(\Phi_2^\dagger \Phi_2\right)^2 + \lambda_3 \left(\Phi_1^\dagger \Phi_1\right)\left(\Phi_2^\dagger \Phi_2\right) \\
& + \lambda_4 \left(\Phi_1^\dagger \Phi_2\right)\left(\Phi_2^\dagger \Phi_1\right) + \frac{\lambda_5}{2} \left[\left(\Phi_1^\dagger \Phi_2\right)^2 + \left(\Phi_2^\dagger \Phi_1\right)^2\right] \\
& + 2 m_S^2 \mathrm{Tr} S^{\dagger i} S_i + \mu_1 \mathrm{Tr} S^{\dagger i} S_i S^{\dagger j} S_j \\
& + \mu_2 \mathrm{Tr} S^{\dagger i} S_j S^{\dagger j} S_i + \mu_3 \mathrm{Tr} S^{\dagger i} S_i \mathrm{Tr} S^{\dagger j} S_j \\
& + \mu_4 \mathrm{Tr} S^{\dagger i} S_j \mathrm{Tr} S^{\dagger j} S_i \\
& + \mu_5 \mathrm{Tr} S_i S_j \mathrm{Tr} S^{\dagger i} S^{\dagger j} + \mu_6 \mathrm{Tr} S_i S_j S^{\dagger j} S^{\dagger i} \\
& + \nu_1 \Phi_1^{\dagger i} \Phi_{1i} \mathrm{Tr} S^{\dagger j} S_j + \nu_2 \Phi_1^{\dagger i} \Phi_{1j} \mathrm{Tr} S^{\dagger j} S_i \\
& + \Big(\nu_3 \Phi_1^{\dagger i} \Phi_1^{\dagger j} \mathrm{Tr} S_i S_j + \nu_4 \Phi_1^{\dagger i} \mathrm{Tr} S^{\dagger j} S_j S_i \\
& + \nu_5 \Phi_1^{\dagger i} \mathrm{Tr} S^{\dagger j} S_i S_j + \text{h.c.}\Big) \\
& + \omega_1 \Phi_2^{\dagger i} \Phi_{2i} \mathrm{Tr} S^{\dagger j} S_j + \omega_2 \Phi_2^{\dagger i} \Phi_{2j} \mathrm{Tr} S^{\dagger j} S_i \\
& + \Big(\omega_3 \Phi_2^{\dagger i} \Phi_2^{\dagger j} \mathrm{Tr} S_i S_j + \omega_4 \Phi_2^{\dagger i} \mathrm{Tr} S^{\dagger j} S_j S_i \\
& + \omega_5 \Phi_2^{\dagger i} \mathrm{Tr} S^{\dagger j} S_i S_j + \text{h.c.}\Big) \\
& + \kappa_1 \Phi_1^{\dagger i} \Phi_{2i} \mathrm{Tr} S^{\dagger j} S_j + \kappa_2 \Phi_1^{\dagger i} \Phi_{2j} \mathrm{Tr} S^{\dagger j} S_i \\
& + \kappa_3 \Phi_1^{\dagger i} \Phi_2^{\dagger j} \mathrm{Tr} S_j S_i + \text{h.c.},
\end{aligned}
\tag{2.1}
$$

where, $\Phi_1$, $\Phi_2$ and $S$ can be written as

$$
\Phi_i = \begin{pmatrix} \phi_i^+ \\ \frac{1}{\sqrt{2}}(v_i + h_i + i z_i) \end{pmatrix}, (i = 1, 2),
$$

$$
S = \begin{pmatrix} S^+ \\ \frac{1}{\sqrt{2}}(S_R + i S_I) \end{pmatrix}. \tag{2.2}
$$

In the above, $v_i$ is the vacuum expectation value (VEV) of $\Phi_i$ with $v^2 = v_1^2 + v_2^2 = (246\,\text{GeV})^2$. The ratio of two VEVs relates to the mixing angle $\beta$ as $\tan\beta = \frac{v_2}{v_1}$. Here the scalar potential parameters $m_{11}^2, m_{22}^2, m_{12}^2, m_S^2, \mu_{1-6}, \lambda_{1-5}, \nu_{1-5}, \omega_{1-5}, k_{1-3}$ are taken real to avoid CP-violation in the scalar sector. Similarly to [9], we are guided by MFV and not by the more restrictive option of a discrete symmetry such as $Z_2$ (under which $\Phi_1 \to \Phi_1,\ \Phi_2 \to -\Phi_2$ for instance).[1]

[1] $\kappa_1, \kappa_2, \kappa_3 \neq 0$ is disallowed by the $Z_2$-symmetry but allowed by MFV. We intend to extract the maximally enhanced contribution coming from the scalar potential to the color-octet form factors and hence, do not commit to a $Z_2$.

In Eq. (2.1), $i, j$ and $A, B$ respectively denote the fundamental $SU(2)$ and adjoint $SU(3)$ indices. One then defines $S_i = S_i^A T^A$ ($T^A$ being the $SU(3)$ generators) and the trace in Eq. (2.1) is taken over the color indices. The colorless particle spectrum in this case is identical with the 2HDM that consists of the neutral CP-even Higgses $h, H$, a CP-odd Higgs $A$ and a charged Higgs $H^+$. The $2 \times 2$ mass matrix corresponding to the CP-even scalars is brought into a diagonal form by the action of a mixing angle $\alpha$. Of these, the scalar $h$ is taken to be the SM-like Higgs with mass 125 GeV. The masses of the neutral and charged mass eigenstate of the color-octet can be written in terms of the quartic couplings $\omega_i, \kappa_i, \nu_i$ and mixing angle $\beta$ as [38]:

$$
\begin{aligned}
M_{S_R}^2 = m_S^2 + \frac{1}{4} v^2 (\cos^2\beta (\nu_1 + \nu_2 + 2\nu_3) \\
+ \sin 2\beta (\kappa_1 + \kappa_2 + \kappa_3) \\
+ \sin^2\beta (\omega_1 + \omega_2 + 2\omega_3)),
\end{aligned}
\tag{2.3a}
$$

$$
\begin{aligned}
M_{S_I}^2 = m_S^2 + \frac{1}{4} v^2 (\cos^2\beta (\nu_1 + \nu_2 - 2\nu_3) \\
+ \sin 2\beta (\kappa_1 + \kappa_2 - \kappa_3) \\
+ \sin^2\beta (\omega_1 + \omega_2 - 2\omega_3)),
\end{aligned}
\tag{2.3b}
$$

$$
M_{S^+}^2 = m_S^2 + \frac{1}{4} v^2 (\nu_1 \cos^2\beta + \kappa_1 \sin 2\beta + \omega_1 \sin^2\beta). \tag{2.3c}
$$

The Yukawa interactions in this model partition into the two following terms:

$$
\mathcal{L}^{\rm Y} = \mathcal{L}_1^Y + \mathcal{L}_8^Y \tag{2.4}
$$

where $\mathcal{L}_{1(8)}^Y$ involves the SM quarks and the color-singlet (octet) electroweak doublet. Here, $\mathcal{L}_1^Y$ is chosen to coincide with the Yukawa Lagrangian in Type-I and Type-II 2HDM that essentially suppress the tree level flavour changing neutral currents (FCNC). On the other hand, the interaction with the color-octet has the form:

$$
\mathcal{L}_8^Y = -(y_u')_{pq} \bar{u}_R^p \tilde{S}^\dagger Q_L^q - (y_d')_{pq} \bar{d}_R^p S^\dagger Q_L^q + \text{h.c.} \tag{2.5}
$$

Here, $(p, q) = 1, 2, 3$ are the fermion generation indices. We however remark that $\mathcal{L}_8^Y$ shall not play a role in the present analysis and we retain it just for completeness.

## 3 One-loop form factors for $H^+ \to W^+ Z(\gamma)$

In this section we compute the $H^+ W^- Z(\gamma)$ vertex at one-loop for the present scenario. The various form factors are expressed in terms of the Passarino–Veltman functions and the publicly available library `LoopTools` [46] is used for numerical evaluation.

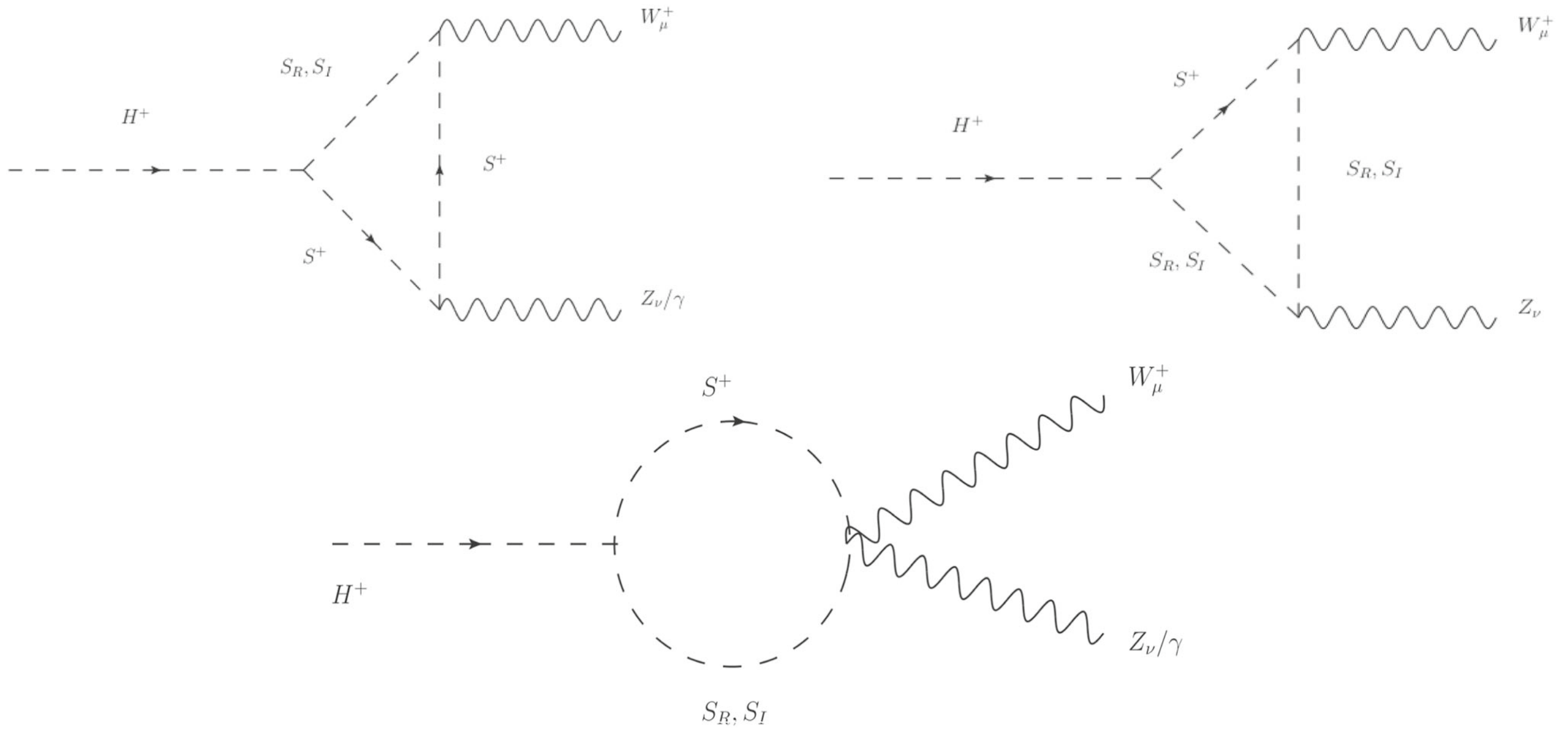

**Fig. 1** Set A of one-loop amplitudes

The amplitude for $H^+ \to W^+V(V = Z, \gamma)$ can be expressed as

$$i\mathcal{M}(H^+ \to W^+V) = igm_W V_V^{\mu\nu} \epsilon^*_{W\mu}(p_W)\epsilon^*_{V\nu}(p_V) \,. \quad (3.1)$$

where,

$$V_V^{\mu\nu} = g^{\mu\nu}F_V + \frac{p_V^\mu p_W^\nu}{m_W^2}G_V + i\epsilon^{\mu\nu\rho\sigma}\frac{p_{V\rho}p_{W\sigma}}{m_W^2}H_V \,. \quad (3.2)$$

Here $p_W^\nu, p_V^\mu$ are the incoming momenta of $W^\pm$ and $V$. Moreover, $F$, $G$ and $H$ are the form-factors corresponding to the respective Lorentz structures. For $V = \gamma$, the Ward identity enforces the following condition:

$$V_\gamma^{\mu\nu} p_{\gamma\nu} = 0 \quad (3.3)$$

This ultimately leads to the following relation connecting $F_\gamma$ and $G_\gamma$:

$$F_\gamma = \frac{G_\gamma}{2}\left(1 - \frac{M_{H^+}^2}{M_W^2}\right) \quad (3.4)$$

Scalars coming from both colorless and colored sectors, i.e, $\Phi_{1,2}$ as well as $S$, contribute to $H^+ \to W^+V$. Accordingly, each form factor splits as

$$X_V = X_{V,2\text{HDM}} + X_{V,S}, \quad (3.5)$$

for $X = F, G, H$. We now come to discussing the various one-loop diagrams comprising the amplitudes. For the colored part, the amplitude receives contributions from the following set of one-loop diagrams shown in Figs. 1, 2 and 3.

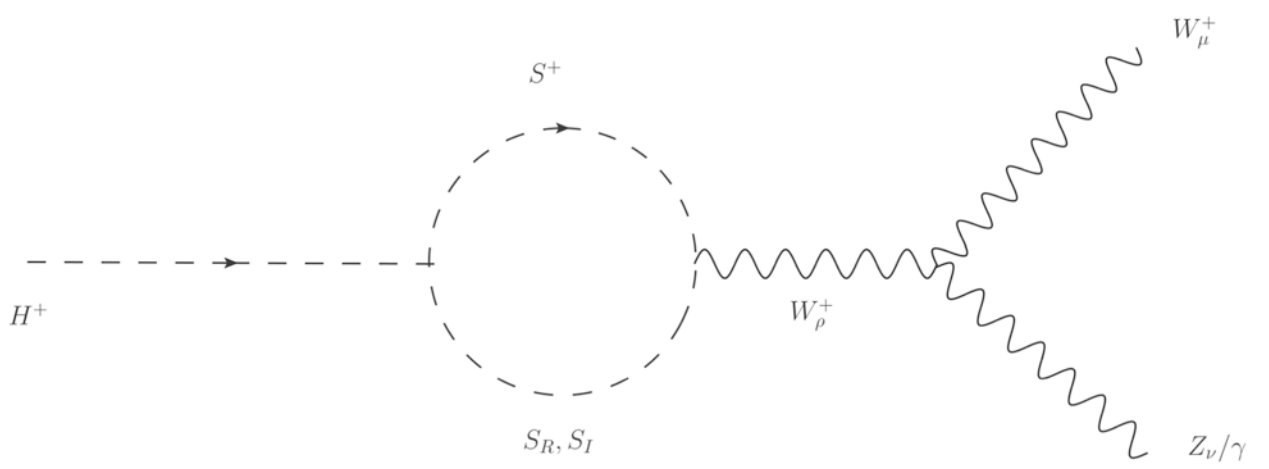

**Fig. 2** Set B of one-loop amplitudes

The amplitude corresponding to each set is UV-finite. To this end we introduce nonlinear gauge-fixing functions [47–53]:

$$f^+ = \left(D_\mu^e + \frac{igs_W^2}{c_W}Z_\mu\right)W^{+\mu} - i\xi M_W G^+, \quad (3.6a)$$

$$f^Z = \partial_\mu Z^\mu - \xi M_Z G^0, \quad (3.6b)$$

$$f^A = \partial_\mu A^\mu. \quad (3.6c)$$

with $D_\mu^e$ the electromagnetic covariant derivative and $\xi$ the gauge parameter. Note that $f^+$ is nonlinear and transforms covariantly under the electromagnetic gauge group. The corresponding gauge fixing Lagrangian is given by

$$\mathcal{L}_{GF} = -\frac{1}{\xi}f^+f^- - \frac{1}{2\xi}(f^Z)^2 - \frac{1}{2\xi}(f^A)^2 \quad (3.7a)$$

This gauge-fixing procedure is tailor-made to remove the unphysical $G^+W^-V$ vertices that arise in the Higgs kinetic-energy pgsector. Further into this, for $\xi = 1$, the $V_\mu(k)W_\nu^+(p)$ $W^-_\rho(q)$ (all momenta incoming) triple-gauge vertices now have the following modified Feynman rules.

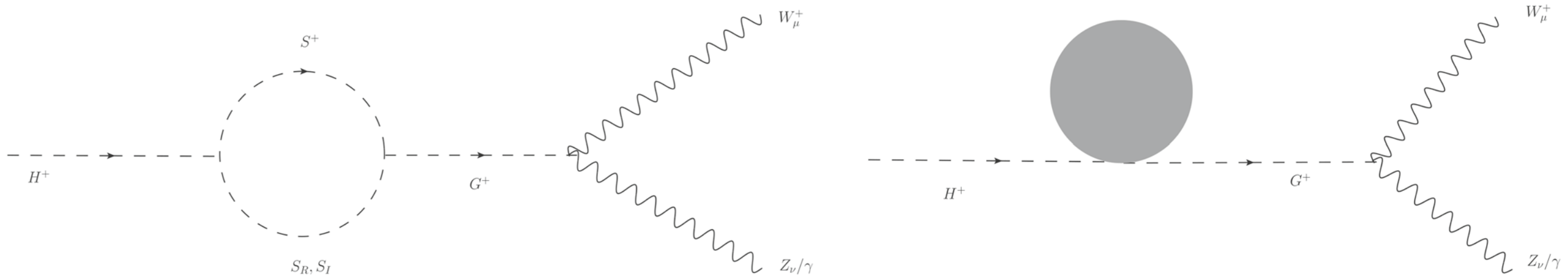

**Fig. 3** Set C of one-loop amplitudes

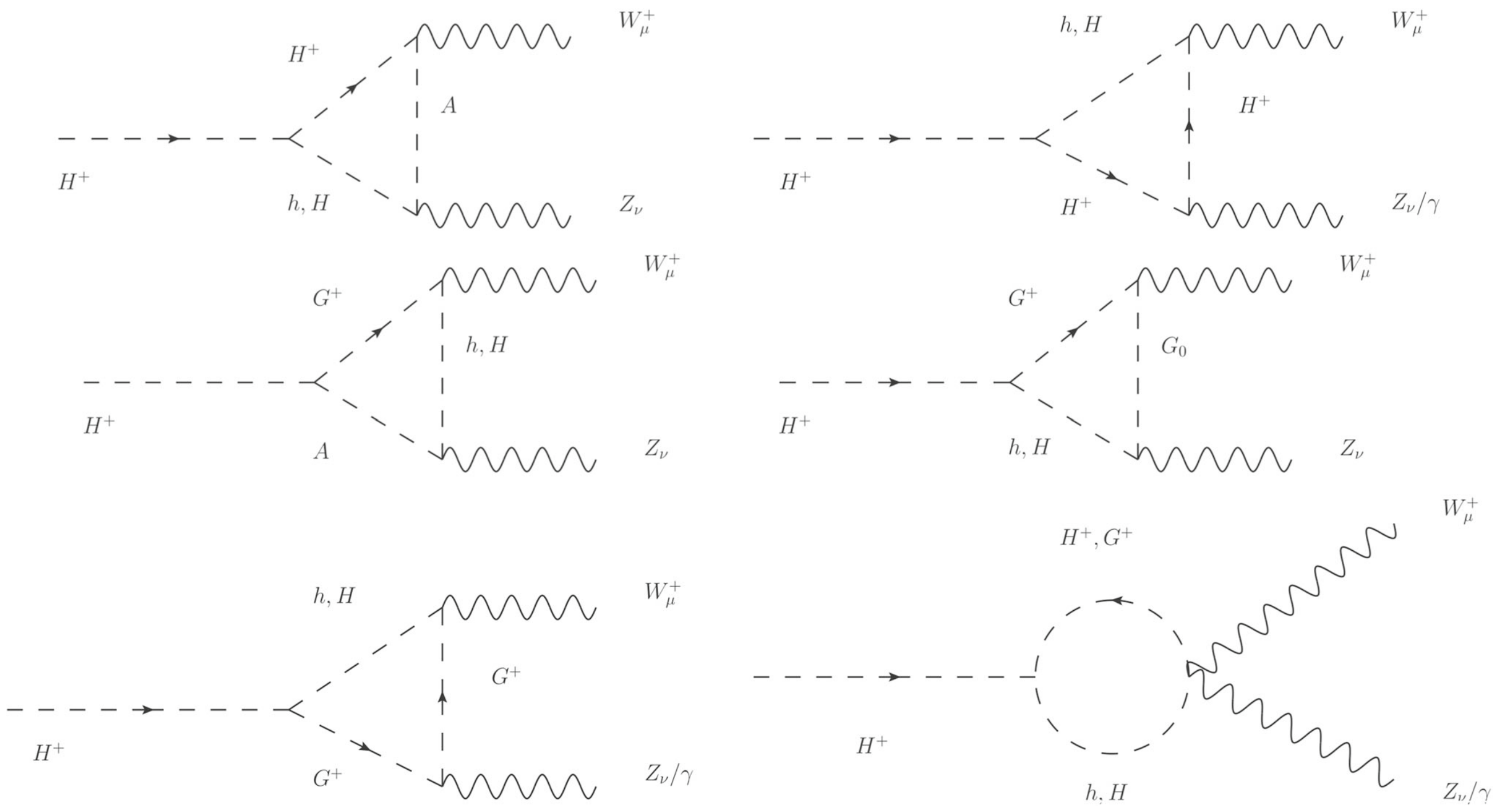

**Fig. 4** Set A′ of one-loop amplitudes for 2HDM

$$\Gamma^{\gamma W^+ W^-}_{\mu\nu\rho}(k,p,q) = -ie\{g_{\mu\nu}(k-p-q)_\rho + g_{\nu\rho}(p-q)_\mu + g_{\rho\mu}(q-k+p)_\nu\}, \tag{3.8a}$$

$$\Gamma^{ZW^+W^-}_{\rho\nu\mu}(k,p,q) = -igc_W\left\{g_{\mu\nu}\left(k-p+\frac{s_W^2}{c_W^2}q\right)_\rho + g_{\nu\rho}(p-q)_\mu + g_{\rho\mu}\left(q-k-\frac{s_W^2}{c_W^2}p\right)_\nu\right\}. \tag{3.8b}$$

Therefore, by virtue of the aforementioned gauge fixing, the amplitudes in Set C (Fig. 3) vanish. Moreover, this implies that similar amplitudes coming from the colorless scalars would also vanish. We partition the remaining 2HDM diagrams into UV-finite sets A′, B′ and C′ as displayed in Figs. 4, 5 and 6.

We point out that Set A′ (Fig. 4) is the color-singlet counterpart of Set A (Fig. 1), both containing a scalar trilinear interaction on one vertex and gauge interactions on the other(s). Set B′ (Fig. 5) can also be related to Set B (Fig. 2) using similar arguments. The additional diagrams in Set C′ (Fig. 6) feature $h(H)VV$ interactions and vanish in the $s_{\beta-\alpha} = 1$ limit. Finally, Fig. 7 contains the fermionic one-loop diagrams.

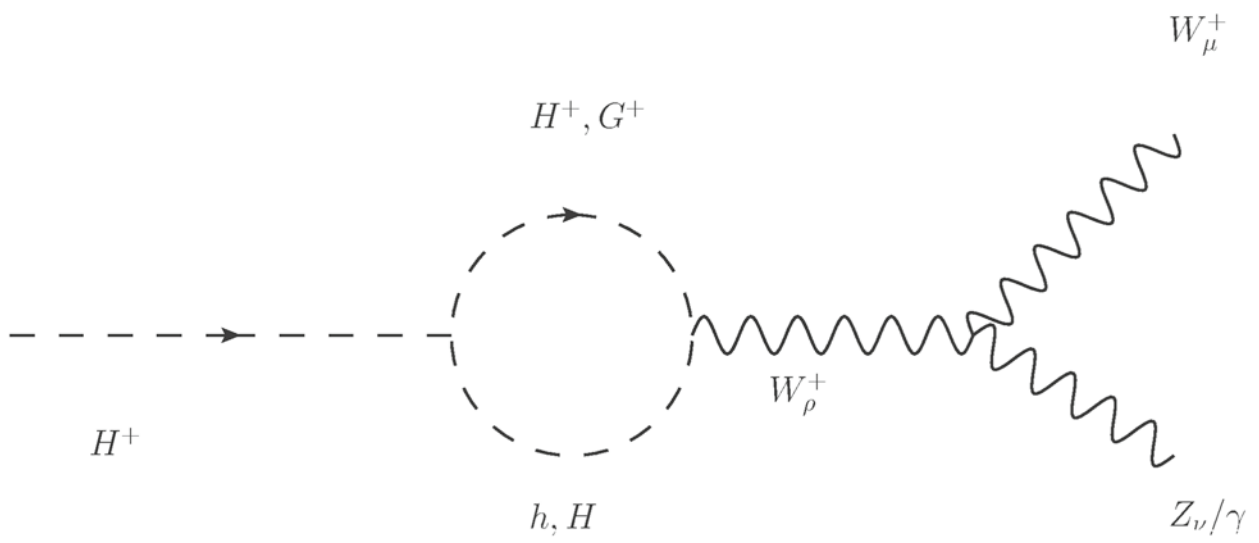

**Fig. 5** Set B′ of one-loop amplitudes for 2HDM

The decay width of $H^+ \to W^+ Z$ is given by

$$\Gamma(H^+ \to W^+ Z) = M_{H^+}\frac{\sqrt{\lambda(1,\omega,z)}}{16\pi}\sum_{i=L,T}|M_{ii}|^2, \tag{3.9}$$

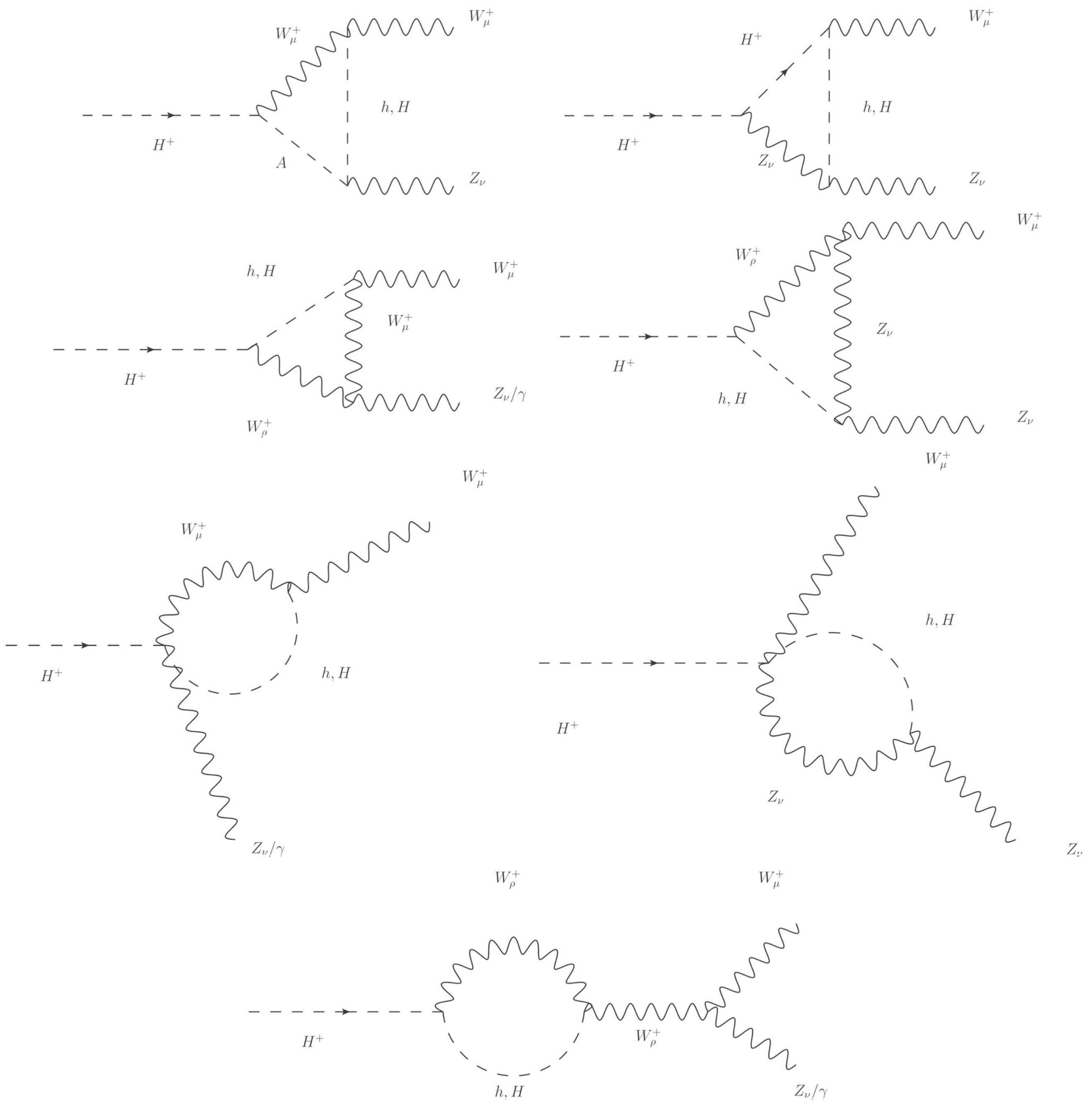

**Fig. 6** Set C′ of one-loop amplitudes for 2HDM

where $i = L(T)$ represents the longitudinal and transverse polarization and, $\lambda(a, b, c) = (a - b - c)^2 - 4abc, \omega = (\frac{M_W}{M_{H^+}})^2, z = (\frac{M_Z}{M_{H^+}})^2$.

The longitudinal and transverse contributions are given in terms of $F_V, G_V, H_V$ by,

$$|M_{LL}|^2 = \frac{g^2}{4z}\left|(1-\omega-z)F_V + \frac{\lambda(1,\omega,z)}{2\omega}G_V\right|^2, \quad (3.10a)$$

$$|M_{TT}|^2 = g^2\left(2\omega|F_V|^2 + \frac{\lambda(1,\omega,z)}{2\omega}|H_V|^2\right). \quad (3.10b)$$

For $V = \gamma$, the relation $F_\gamma = \frac{G_\gamma}{2}\left(1 - \frac{M_{H^+}^2}{M_W^2}\right)$ is used in Eqs. (3.10a) and (3.10b) to obtain

$$\Gamma_{H^+\to W^+\gamma} = \frac{M_{H^+}^3}{8\pi v^2}\left(1 - \frac{M_{W^+}^2}{M_{H^+}^2}\right)^3 (|G_\gamma|^2 + |H_\gamma|^2). \quad (3.11)$$

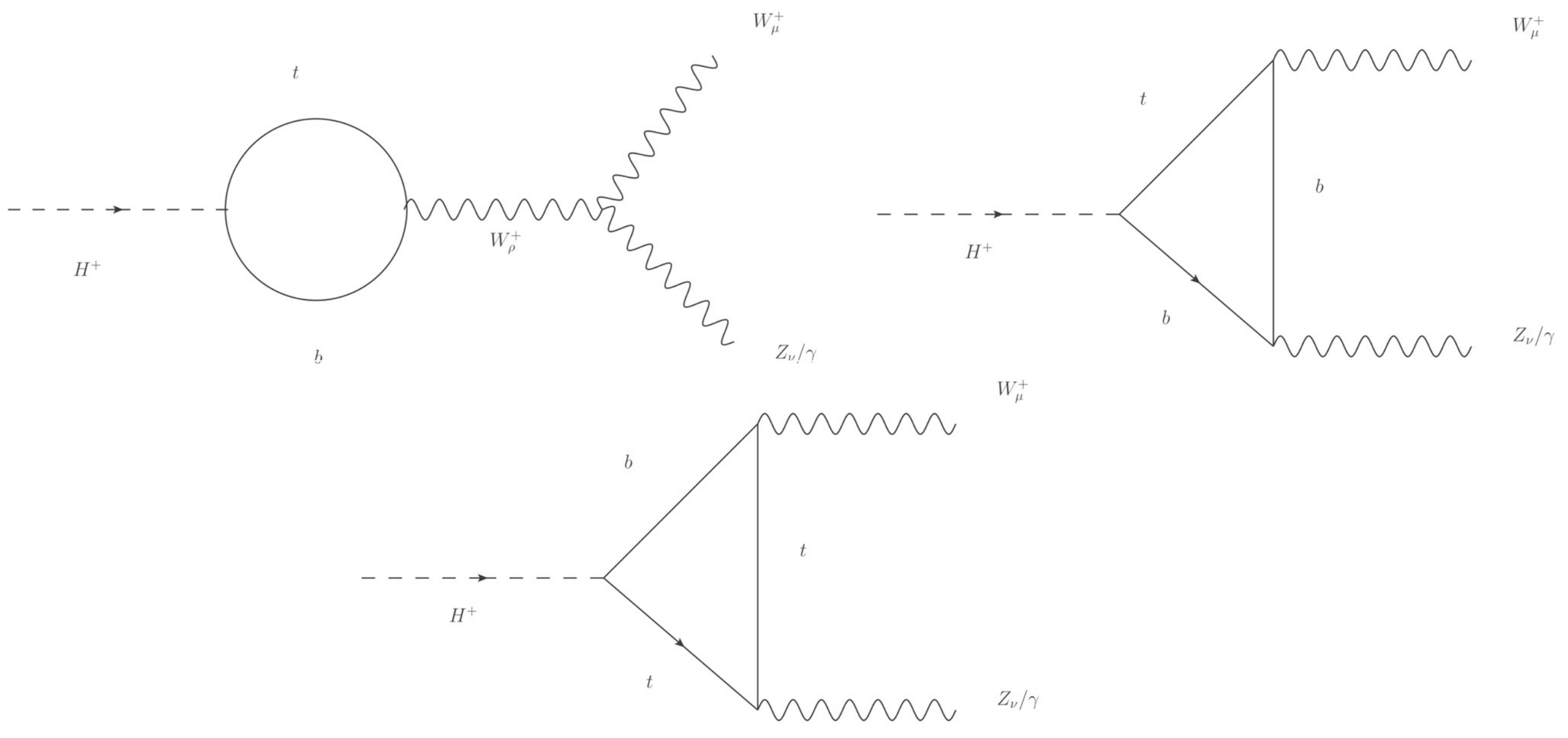

**Fig. 7** Fermionic one-loop amplitudes

## 4 Constraints applied

We discuss the relevant constraints on this model in this section.

### 4.1 Perturbativity

Demanding perturbativity leads to the following bounds on the couplings:

$$|\lambda_i| \leq 4\pi \ (i = 1, 2, ...5), \ |\nu_j|, |\omega_j|, |k_j| \leq 4\pi \ (j = 1, 2, 3), |y_i| \leq 4\pi \ (i = t, b, \tau). \quad (4.1a)$$

### 4.2 Stability conditions

To make the scalar potential bounded-from-below (BFB) in any direction in the field space, the following conditions are to be satisfied [54]:

$$\mu = \mu_1 + \mu_2 + \mu_6 + 2(\mu_3 + \mu_4 + \mu_5) > 0, \quad (4.2a)$$
$$\mu_1 + \mu_2 + \mu_3 + \mu_4 > 0, \quad (4.2b)$$
$$14(\mu_1 + \mu_2) + 5\mu_6 + 24(\mu_3 + \mu_4) - 3|2(\mu_1 + \mu_2) - \mu_6| > 0, \quad (4.2c)$$
$$5(\mu_1 + \mu_2 + \mu_6) + 6(2\mu_3 + \mu_4 + \mu_5) - |\mu_1 + \mu_2 + \mu_6| > 0, \quad (4.2d)$$
$$\lambda_1 \geq 0, \ \lambda_2 \geq 0, \ \lambda_3 \geq -\sqrt{\lambda_1\lambda_2}, \quad (4.2e)$$
$$\lambda_3 + \lambda_4 - |\lambda_5| \geq -\sqrt{\lambda_1\lambda_2}, \quad (4.2f)$$
$$\nu_1 \geq -2\sqrt{\lambda_1\mu}, \quad (4.2g)$$
$$\omega_1 \geq -2\sqrt{\lambda_2\mu}, \quad (4.2h)$$
$$\nu_1 + \nu_2 - 2|\nu_3| \geq -2\sqrt{\lambda_1\mu}, \quad (4.2i)$$
$$\omega_1 + \omega_2 - 2|\omega_3| \geq -2\sqrt{\lambda_2\mu}, \quad (4.2j)$$
$$\lambda_1 + \frac{\mu}{4} + \nu_1 + \nu_2 + 2\nu_3 - \frac{1}{\sqrt{3}}|\nu_4 + \nu_5| > 0, \quad (4.2k)$$
$$\lambda_2 + \frac{\mu}{4} + \omega_1 + \omega_2 + 2\omega_3 - \frac{1}{\sqrt{3}}|\omega_4 + \omega_5| > 0. \quad (4.2l)$$

Among the above, Eqs. (4.2e) and (4.2f) correspond to the pure 2HDM. The rest of the conditions ensure positivity of the scalar potential in a hyperspace spanned by both colorless as well as colored fields. It is pointed out that Eqs. (4.2a)–(4.2d) involve only the parameters $\mu_{1-6}$ that do not enter our analysis.

### 4.3 High energy scattering unitarity

Additional constraints on the quartic couplings come from the unitarity. A tree-level $2 \to 2$ scattering matrix can be computed between various two particle states consisting of charged and neutral scalars [55]. Unitarity requires that the absolute value of each eigenvalue of the aforementioned scattering matrix must be $\leq 8\pi$. In the high energy limit, one such element of the scattering matrix is proportional to a quartic coupling. Therefore, demanding unitarity is tantamount to restricting the sizes of such couplings. Thus we end up with the following unitarity conditions for the present model [38].

$$\left[\frac{3}{2}(\lambda_1 + \lambda_2) \pm \sqrt{\frac{9}{4}(\lambda_1 - \lambda_2)^2 + (2\lambda_3 + \lambda_4)^2}\right] \leq 8\pi, \quad (4.3a)$$

$$\left[\frac{1}{2}(\lambda_1+\lambda_2) \pm \sqrt{\frac{1}{4}(\lambda_1-\lambda_2)^2+\lambda_4^2}\right] \leq 8\pi, \tag{4.3b}$$

$$\left[\frac{1}{2}(\lambda_1+\lambda_2) \pm \sqrt{\frac{1}{4}(\lambda_1-\lambda_2)^2+\lambda_5^2}\right] \leq 8\pi, \tag{4.3c}$$

$$(\lambda_3+2\lambda_4-3\lambda_5) \leq 8\pi, \tag{4.3d}$$

$$(\lambda_3-\lambda_5) \leq 8\pi, \tag{4.3e}$$

$$(\lambda_3+\lambda_4) \leq 8\pi, \tag{4.3f}$$

$$(\lambda_3+2\lambda_4+3\lambda_5) \leq 8\pi, \tag{4.3g}$$

$$(\lambda_3+\lambda_5) \leq 8\pi, \tag{4.3h}$$

$$|\nu_1| \leq 2\sqrt{2}\pi,\ |\nu_2| \leq 4\sqrt{2}\pi,\ |\nu_3| \leq 2\sqrt{2}\pi, \tag{4.3i}$$

$$|\omega_1| \leq 2\sqrt{2}\pi,\ |\omega_2| \leq 4\sqrt{2}\pi,\ |\omega_3| \leq 2\sqrt{2}\pi, \tag{4.3j}$$

$$|\kappa_1| \leq 2\pi,\ |\kappa_2| \leq 4\pi,\ |\kappa_3| \leq 4\pi, \tag{4.3k}$$

$$|17\mu_3+13\mu_4+13\mu_6| \leq 16\pi, \tag{4.3l}$$

$$|2\mu_3+10\mu_4+7\mu_6| \leq 32\pi, \tag{4.3m}$$

$$|\nu_4+\nu_5| \lesssim \frac{32\pi}{\sqrt{15}}, \tag{4.3n}$$

$$|\omega_4+\omega_5| \lesssim \frac{32\pi}{\sqrt{15}}. \tag{4.3o}$$

Equations (4.3a)–(4.3h) correspond to the unitarity limit for a pure two-Higgs doublet scenario [56–62]. Once again, we state that Eqs. (4.3l)–(4.3o) involve $\mu_{1-6}, \omega_{4,5}, \nu_{4,5}$ alone. These conditions therefore do not modify the analysis. More details on unitarising a spinless (8, 2, 1/2) multiplet can be found in [29,38].

## 4.4 Oblique parameters

The extended scalar sector modifies the oblique parameters [63] with respect to the SM contributions. The strongest constraint however comes from the $T$-parameter. Including the BSM contribution, the effective $T$-parameter can be written in terms of the SM and BSM contribution $\Delta T$ as:

$$T = T_{\rm SM} + \Delta T\ . \tag{4.4}$$

The most updated bound on the BSM contribution to $T$-parameter is [64]:

$$\Delta T = 0.07 \pm 0.12. \tag{4.5}$$

For the multi-Higgs doublet models, The most constraining among the oblique parameters is the $T$-parameter [65] that restricts the mass splittings between the neutral and charged scalars. In our case, the source of BSM contribution is two-fold, i.e., contribution arising from 2HDM and the scalar octet. Thus,

$$\Delta T = T_{\rm 2HDM} + T_S, \tag{4.6a}$$

$$\begin{aligned} T_{\rm 2HDM} = \frac{1}{16\pi s_W^2 M_W^2}\Big[&F(M_{H^+}^2, M_A^2) + s_{\beta-\alpha}^2 \\ &\times \Big(F(M_{H^+}^2, M_H^2) - F(M_H^2, M_A^2)\Big) \\ &+ c_{\beta-\alpha}^2\Big(F(M_{H^+}^2, M_h^2) - F(M_A^2, M_h^2) \\ &+ F(M_W^2, M_H^2) - F(M_W^2, M_h^2) \\ &+ F(M_Z^2, M_h^2) - F(M_Z^2, M_H^2) \\ &+ 4M_Z^2\bar{B}_0(M_Z^2, M_H^2, M_h^2) - 4M_W^2\bar{B}_0(M_W^2, M_H^2, M_h^2)\Big)\Big], \end{aligned}$$

$$T_S = \frac{N_S}{16\pi s_W^2 M_W^2} \times \Big[F(M_{S^+}^2, M_{S_R}^2) + F(M_{S^+}^2, M_{S_I}^2) - F(M_{S_R}^2, M_{S_I}^2)\Big], \tag{4.6b}$$

where,

$$F(x,y) = \frac{x+y}{2} - \frac{xy}{x-y}\ln\left(\frac{x}{y}\right) \quad \text{for} \quad x \neq y,$$
$$= 0 \quad \text{for} \quad x = y \tag{4.7a}$$

$$\bar{B}_0(m_1^2, m_2^2, m_3^2) = \frac{m_1^2\log(m_1^2) - m_3^2\log(m_3^2)}{m_1^2 - m_3^2} - \frac{m_1^2\log(m_1^2) - m_2^2\log(m_2^2)}{m_1^2 - m_2^2}. \tag{4.7b}$$

## 4.5 Flavour constraints

For a 2HDM with natural flavour conservation, the strongest constraint on the mass of the charged Higgs comes from the branching fraction for $B \to X_s\gamma$. In case of the Type-I 2HDM, this particular constraint supersedes the direct search constraints only for small tan$\beta$. On the other hand, the Type-II 2HDM features a stringent $M_{H^+} > 580$ GeV bound [66] that is practically independent of tan$\beta$.

## 4.6 Direct search

An $M_{H^+} > 100$ GeV bound for all types of 2HDM summarises the result for the direct search of $H^+$ in the $e^+e^- \longrightarrow H^+H^-$ channel at LEP [67]. In the $M_{H^+} > M_t + M_b$ region, a measurement of $\sigma(pp \longrightarrow \bar{t}H^+ + X) \times {\rm BR}(H^+ \longrightarrow \bar{\tau}\nu_\tau)$ obviates tan$\beta > 50$ for an $H^+$ of mass $\simeq 200$ GeV in case of Type-II 2HDM [68]. The corresponding constraint is further weakened for Type-I 2HDM. In view of the aforementioned, we take $M_{H^+} \geq 150\,(600)$ GeV for a Type-I (Type-II) 2HDM throughout the present study.

We now come to discussing the exclusion limits on the color-octet mass scale itself. Color-octet resonances have been searched for at the LHC in the $pp \to S \to jj$ [69–72] and $pp \to S \to t\bar{t}$ [73–75] channels and a stringent bound of $\gtrsim 3$ TeV was subsequently pronounced. However, it was pointed out in [76] that the benchmark color-octet taken there led to a cross section several orders larger than what would be seen for the Manohar–Wise scenario. Ref.[76] showed that the bound weakens to $M_{S_R} > 700$ GeV upon tweaking the model parameters appropriately. On the other hand, pair-production of $S$, that occurs at the tree level itself, yields a cross section comparable to the loop-induced sin-gle $S$ production. The pair-production was also studied at the LHC in the $4j$, $4b$, $4t$, $t\bar{t}b\bar{b}$ final states. Ref. [76] showed that $M_{S_R} \simeq 800$ GeV can evade all such constraints.

**Table 1** Latest limits on the $h$-signal strengths

| $\mu_i$ | ATLAS | CMS |
|---|---|---|
| $ZZ$ | $1.20^{+0.16}_{-0.15}$ [77] | $0.94^{+0.07}_{-0.07}$(stat.)$^{+0.08}_{-0.07}$(syst.) [78] |
| $W^+W^-$ | $2.5^{+0.9}_{-0.8}$ [79] | $1.28^{+0.18}_{-0.17}$ [80] |
| $\gamma\gamma$ | $0.99 \pm 0.14$ [81] | $1.18^{+0.17}_{-0.14}$ [82] |
| $\tau\bar{\tau}$ | $1.09^{+0.18}_{-0.17}$(stat.)$^{+0.27}_{-0.22}$(syst)$^{+0.16}_{-0.11}$(theo syst) [83] | $1.09^{+0.27}_{-0.26}$ [84] |
| $b\bar{b}$ | $2.5^{+1.4}_{-1.3}$ [85] | $1.3^{+1.2}_{-1.1}$ [86] |

### 4.7 Higgs signal strengths

The model parameters are also constrained from the current LHC data, i.e. signal strengths in various Higgs decay decay modes. Denoting the signal strength for the channel $pp \to h,\ h \to i$ by $\mu_i$, it is defined as,

$$\mu_i = \frac{\sigma^{\rm BSM}(pp \to h)\ {\rm BR}^{\rm BSM}(h \to i)}{\sigma^{\rm SM}(pp \to h)\ {\rm BR}^{\rm SM}(h \to i)}. \tag{4.8}$$

Equation (4.8) takes the form below upon expressing the branching fractions in terms of the decay widths.

$$\mu_i = \frac{\sigma^{\rm BSM}(gg \to h)}{\sigma^{\rm SM}(gg \to h)} \frac{\Gamma_i^{\rm BSM}(h \to i)}{\Gamma_{\rm tot}^{\rm BSM}} \frac{\Gamma_{\rm tot}^{\rm SM}}{\Gamma_i^{\rm SM}(h \to i)}. \tag{4.9}$$

Note that here the BSM contribution includes contributions from both 2HDM and the color-octet. In particular, the modification induced by the color-octet is at the leading order (LO) is solely to the loop-induced $h \to gg,\ \gamma\gamma,\ Z\gamma$ decays.[2]

Now the parton-level cross section of Higgs production through gluon fusion can be written as

$$\sigma(gg \to h) = \frac{\pi^2}{8M_h}\Gamma(h \to gg)\ \delta(\hat{s} - M_h^2), \tag{4.10}$$

$\sqrt{\hat{s}}$ being partonic centre-of-mass energy. Using Eqs. (4.8)–(4.10), one can rewrite the $\mu_i$ as :

$$\mu_i = \frac{\Gamma^{\rm BSM}_{h\to gg}}{\Gamma^{\rm SM}_{h\to gg}} \frac{\Gamma_i^{\rm BSM}}{\Gamma_{\rm tot}^{\rm BSM}} \frac{\Gamma_{\rm tot}^{\rm SM}}{\Gamma_i^{\rm SM}}. \tag{4.11}$$

[2] Since $h \to WW,\ ZZ,\ bb$ decays occur at the tree level itself, one-loop corrections to these modes from the color-octet are at the next-to-leading order (NLO) level. All of these one-loop amplitudes involve the couplings of $h$ to a pair of color-octet scalars. Such couplings already get constrained from $h \to gg,\ \gamma\gamma$ in particular. Therefore, the NLO signal strengths too are expected to obey the corresponding limits. In addition, color-octet corrections to $h \to \tau\tau$ appear at two-loop, or equivalently, at the next-to-next-to-leading order (NNLO) level. Such corrections are expected to be considerably suppressed.

In presence of color-octet, both the production cross section of Higgs boson via gluon fusion and $h \to gg$ decay width (Eq. (A.4b)) get modified. We have strictly imposed the *alignment limit i.e.* $(\beta - \alpha) = \frac{\pi}{2}$ throughout the analysis. We then compute the signal strengths in the tree level decay modes of Higgs, i.e. $WW$, $ZZ$, $b\bar{b}$, $\tau^+\tau^-$, following Eq. (4.11). The decay width and hence the signal strength in the loop induced decay channel of Higgs $h \to \gamma\gamma$ are also altered owing to the presence of extra charged particles $(H^\pm, S^\pm)$ appearing in the loop. The expressions for the respective decay widths in the loop induced decay modes of $h$, i.e. $h \to \gamma\gamma$, $h \to Z\gamma$, $h \to gg$, are relegated to Appendix A.1 (Table 1).

The signal strength data from the ATLAS and CMS for a given channel can be combined to yield a resultant central value $\mu$ and a resultant 1-sigma uncertainty $\sigma$ as $\frac{1}{\sigma^2} = \frac{1}{\sigma^2_{\rm ATLAS}} + \frac{1}{\sigma^2_{\rm CMS}}$ and $\frac{\mu}{\sigma^2} = \frac{\mu_{\rm ATLAS}}{\sigma^2_{\rm ATLAS}} + \frac{\mu_{\rm CMS}}{\sigma^2_{\rm CMS}}$. In addition, we also require ${\rm BR}_{h\to Z\gamma} < 1\%$ [87].

## 5 Numerical results

We numerically evaluate the $H^+W^-Z(\gamma)$ form factors and the corresponding branching ratios in this section taking Type-I and Type-II-like Yukawa interactions. Prior to that, we list out the independent parameters in this model. The 2HDM sector comprises of tan$\beta$, the quartic couplings $\lambda_{1-5}$ and $m_{12}$. Of these, $\lambda_{1-5}$ can be traded off for the masses of the physical scalars and mixing angles $\alpha$, $\beta$ using the following formulae:

$$\lambda_1 = \frac{M_H^2 c_\alpha^2 + M_h^2 s_\alpha^2 - m_{12}^2 t_\beta}{v^2 c_\beta^2}, \tag{5.1a}$$

$$\lambda_2 = \frac{M_H^2 s_\alpha^2 + M_h^2 c_\alpha^2 - m_{12}^2/t_\beta}{v^2 s_\beta^2}, \tag{5.1b}$$

$$\lambda_3 = \frac{2M_{H^+}^2}{v^2} + \frac{s_{2\alpha}}{s_{2\beta}}\left(\frac{M_H^2 - M_h^2}{v^2}\right) - \frac{m_{12}^2}{v^2 s_\beta c_\beta}, \tag{5.1c}$$

$$\lambda_4 = \frac{M_A^2 - 2M_{H^+}^2}{v^2} + \frac{m_{12}^2}{v^2 s_\beta c_\beta}, \tag{5.1d}$$

$$\lambda_5 = \frac{m_{12}^2}{v^2 s_\beta c_\beta} - \frac{M_A^2}{v^2}. \tag{5.1e}$$

The independent parameters coming from the 2HDM sector are therefore taken to be $(M_h, M_H, M_A, M_{H^+}, t_\beta)$. Similarly, the color-octet sector comprises the mass parameter $M_S$ and the quartic couplings $\mu_{1-6}, \nu_{1-3}, \omega_{1-3}, \kappa_{1-3}$. We can exclude $\mu_{1-6}$ here since they do not enter the current analysis. We again choose to trade off a few quartic couplings in terms of the masses of the colored scalars using:

$$\omega_2 = \frac{2(M_{S_R}^2 + M_{S_I}^2 - 2M_{S^+}^2) - \nu_2 v^2 c_\beta^2 - \kappa_2 v^2 s_{2\beta}}{v^2 s_\beta^2}, \tag{5.2a}$$

$$\omega_3 = \frac{M_{S_R}^2 - M_{S_I}^2 - \nu_3 v^2 c_\beta^2 - \kappa_2 v^2 s_\beta c_\beta}{v^2 s_\beta^2}. \tag{5.2b}$$

The independent parameters in the color-octet sector are therefore $(\nu_1, \nu_2, \nu_3, \kappa_1, \kappa_2, \kappa_3, \omega_1, M_{S_R}, M_{S_I}, M_{S^+})$. We further set $M_{S^+} = M_{S_R}$ and $M_{H^+} = M_H$ throughout the calculation since this leads to a vanishing contribution to the $T$-parameter from the colored scalars (see Eq. (4.6b)) and a manageable contribution from the 2HDM sector (see Eq. (4.6b)). We take $M_{S_R}$ = 800 GeV for this study.

To extract more insight, we derive simplified expressions for the color-octet form factors in the $M_{S_R} \to \infty$ limit[3] as shown below.

$$F_{Z,S}^A = \frac{N_S}{16\pi^2 v c_W}[\lambda_{H^+S^-S_R}\, f_1(r) + \lambda_{H^+S^-S_I}\, f_2(r)] \tag{5.3a}$$

$$F_{Z,S}^B = \frac{N_S c_W}{16\pi^2 v}[\lambda_{H^+S^-S_R} f_3(r) + \lambda_{H^+S^-S_I} f_4(r)] \tag{5.3b}$$

$$G_{Z,S}^A = \frac{N_S M_W^2}{16\pi^2 v c_W M_{S_R}^2}[\lambda_{H^+S^-S_R}\, g_1(r) + \lambda_{H^+S^-S_I}\, g_2(r)] \tag{5.3c}$$

$$F_{\gamma,S}^A = \frac{N_S s_W}{16\pi^2 v}[\lambda_{H^+S^-S_R}\, f_5(r) + \lambda_{H^+S^-S_I}\, f_6(r)] \tag{5.3d}$$

$$F_{\gamma,S}^B = \frac{N_S s_W}{16\pi^2 v}[\lambda_{H^+S^-S_R} f_3(r) + \lambda_{H^+S^-S_I} f_4(r)] \tag{5.3e}$$

$$G_{\gamma,S}^A = -\frac{4N_S M_W^2 s_W}{16\pi^2 v M_{S_R}^2}[\lambda_{H^+S^-S_R}\, g_3(r) + \lambda_{H^+S^-S_I}\, g_4(r)] \tag{5.3f}$$

Here, $r = \frac{M_{S_I}^2}{M_{S_R}^2}$ and $N_S = 8$ (color factor). The functions $f_i(r)$ and $g_i(r)$ have the forms:

$$f_1(r) = \frac{2r^2 \ln r + (4-3r)r - 1}{4(r-1)^2}, \tag{5.4a}$$

$$f_2(r) = \frac{1}{4}\left[4\sin^2\theta_W \left\{\frac{r + r(-\ln r) - 1}{r-1}\right\} + \frac{\cos 2\theta_W\{(r-1)(-2(r-1)\ln r + 3r - 1) + (2-4r)\ln r\}}{(r-1)^2} + 2\ln(2-r) - \frac{r^2 + 2\ln(2-r) - 1}{(r-1)^2}\right], \tag{5.4b}$$

$$f_3(r) = 0, \tag{5.4c}$$

$$f_4(r) = \frac{r^2 - 2r\ln r - 1}{2(r-1)^2}, \tag{5.4d}$$

$$g_1(r) = \frac{3(r-1)^4\cos 2\theta_W + 2(r-1)\{r(11\,r-7) + 18(r-1)^2\ln\left(\frac{1}{r}\right) + 2\}}{36(r-1)^4} + \frac{12\{(r-3)r(2r-3) - 3\}\ln r}{36(r-1)^4}, \tag{5.4e}$$

$$g_2(r) = \frac{r^3 + \cos 2\theta_W\left(2r^3 + 3r^2 - 6r^2\ln r - 6r + 1\right) - 9r + 6(r-2)\ln(2-r) + 8}{6(r-1)^4}, \tag{5.4f}$$

$$f_5(r) = 0, \tag{5.4g}$$

$$f_6(r) = -f_4(r), \tag{5.4h}$$

$$g_3(r) = \frac{1}{24}, \tag{5.4i}$$

$$g_4(r) = \frac{(r-1)\{r(2r+5) - 1\} - 6r^2\ln r}{12(r-1)^4}. \tag{5.4j}$$

[3] This is equivalent to taking $M_{S_R} >> M_{h,H,A,H^+}$. All the EW scale masses in the loop integrals are therefore neglected in this limit.

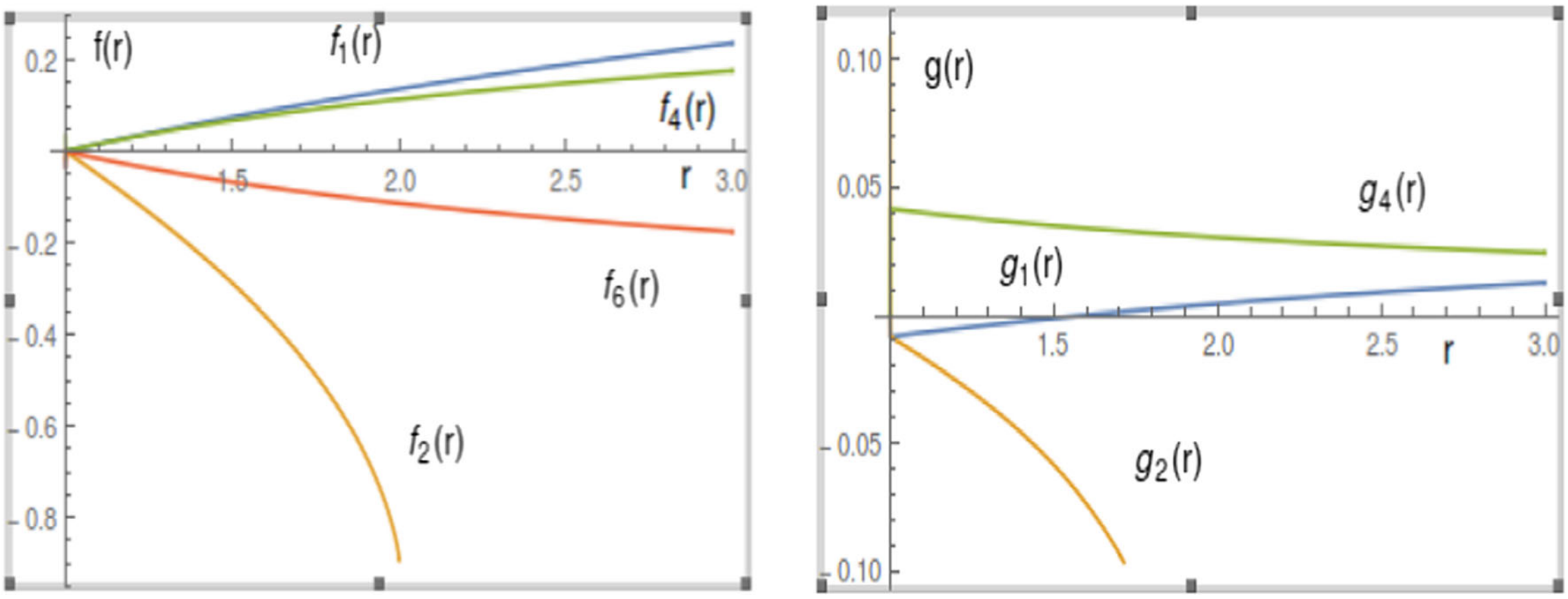

**Fig. 8** Variation of $f_i(r)$ and $g_i(r)$ with $r$

In the above, we have denoted a trilinear interaction involving the arbitrary scalars $S_1$, $S_2$ and $S_3$ (all incoming) as $\lambda_{S_1 S_2 S_3}$. We define $F_{Z(\gamma),S} = F^A_{Z(\gamma),S} + F^B_{Z(\gamma),S}$ and $G_{Z(\gamma),S} = G^A_{Z(\gamma),S}$. It is seen that $G_{Z(\gamma),S} \to 0$ as $M_{S_R} \to \infty$. This is only expected since $G_{Z(\gamma),S}$ encodes the strength of the dimension-6 operator connecting $H^+$, $W^+$ and $Z(\gamma)$ that naturally *decouples* from the *low-energy* theory as the color-octet mass scale becomes too large. Moreover, one also notes $F^A_{\gamma,S} + F^B_{\gamma,S} \simeq \frac{N_S s_W}{16\pi^2 v}\lambda_{H^+S^-S_I}\{f_4(r) + f_6(r)\} = 0$ in the $M_{S_R} \to \infty$ limit. This vanishing of $F_{\gamma,S}$ is anyway expected from Ward identity that reads $F_{\gamma,S} = \frac{G_{\gamma,S}}{2}\left(1 - \frac{M^2_{H^+}}{M^2_W}\right)$.[4] In other words, a vanishing $G_{\gamma,S}$ for $M_{S_R} \to \infty$ implies a vanishing $F_{\gamma,S}$. Therefore, this entire exercise involving the approximate form factors serves as an analytical demonstration of the Ward identity. We hence conclude that the contribution coming from the color-octet sector in the $H^+ \to W^+\gamma$ decay width decouples for a very heavy color-octet mass scale. On the other hand, Eqs. (5.3a) and (5.3b) ascertain that $F^S_Z$ does not exhibit a decoupling behaviour, the most crucial difference between the $H^+W^-Z$ and $H^+W^-\gamma$ form factors at one-loop. We plot $f_i(r)$ and $g_i(r)$ in Fig. 8 and find smooth variations. All $|f_i(r)|$ register increments with increasing $\frac{M^2_{S_I}}{M^2_{S_R}}$, and, therefore $M^2_{S_I} - M^2_{S_R}$ has a role in determining the size of the $H^+W^-Z(\gamma)$ interaction. The trilinear interactions entering the color-octet form factors, i.e., $\lambda_{H^+S^-S_R}$ and $\lambda_{H^+S^-S_I}$, have the following forms for $M_{S^+} = M_{S_R}$.

$$\lambda_{H^+S^-S_R} = -\frac{v}{4}\big(k_2 + k_3 + (\nu_2 + 2\nu_3)\cot\beta\big) \tag{5.5a}$$

$$\lambda_{H^+S^-S_I} = \frac{M^2_{S_I} - M^2_{S_R}}{v} - \frac{v}{4}\big(k_2 - k_3 + (\nu_2 - 2\nu_3)\cot\beta\big) \tag{5.5b}$$

It is read from the equations above that $\lambda_{H^+S^-S_{R/I}}$ are non-zero only in presence electroweak (EW) symmetry breaking.[5] A simple way to see this is that in case of an intact EW symmetry, $M_{S_R} = M_{S_I} = m_S$ and $v = 0$ in which case such $\lambda_{H^+S^-S_{R/I}}$ naturally vanish. An inspection of Fig. 8 and Eq. (5.5b) shows that both the loop functions and the trilinear interactions are sensitive to the mass-splitting between the CP-even and -odd members of the color-octet. To estimate the maximum splitting allowed, we scan $\tan\beta$ and the color-octet parameters in the following ranges:

$$\begin{aligned}
&|\nu_1|, |\nu_2|, |\nu_3|, |\kappa_1|, |\kappa_2|, |\kappa_3|, |\omega_1| \le 4\pi,\\
&2 < \tan\beta < 10,\\
&M_{S_R} < M_{S_I} < M_{S_R} + 500 \text{ GeV}.
\end{aligned} \tag{5.6}$$

The maximum allowed value of $(M_{S_I} - M_{S_R})$ for $M_{S_R} = 800$ GeV allowed by all the constraints is $\simeq 150$ GeV for any $\tan\beta$ as depicted by the left panel of Fig. 9. A scan of the color-octet input parameters for $(M_{S_I} - M_{S_R})$ fixed to 50 GeV, 100 GeV and 150 GeV reveals that the higher the mass splitting, the larger $F_{Z,S}$ can become. This is precisely what the right panel of Fig. 9 shows.

With an insight now gained on the behaviour of the loop functions and the trilinear couplings, we embark on computing the form factors. The following values for the SM parameters are taken: $M_W = 80.3$ GeV, $M_Z = 91.2$ GeV, $M_h = 125.0$ GeV, $M_t = 163$ GeV [88], $M_b = 2.7$ GeV [89],

4 The colorless and color-octet form factors must satisfy the Ward identity separately.

5 A gauge invariant cubic term involving a colorless isodoublet and two color-octet isodoublets does not exist.

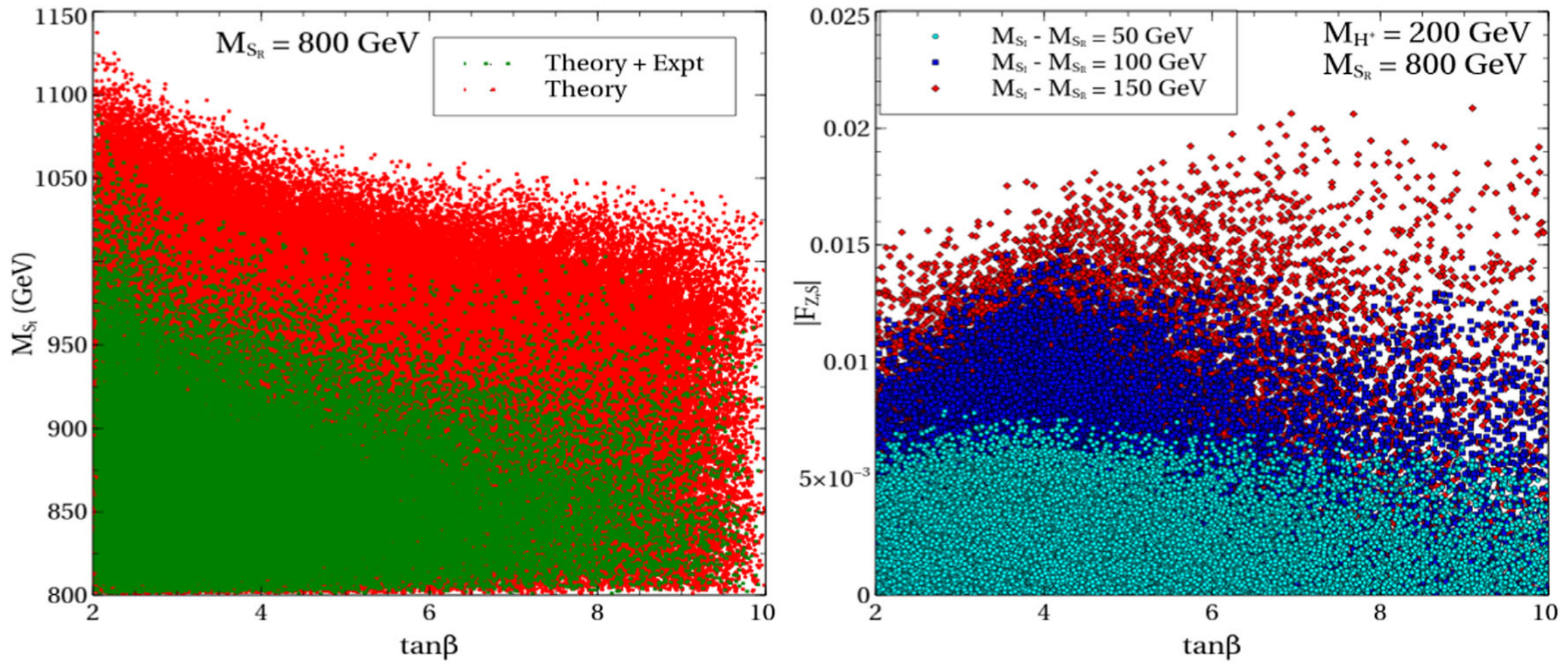

**Fig. 9** Left panel: Allowed parameter points in the $M_{S_I}$-tan$\beta$ plane for $M_{S_R} = M_{S^+} = 800$ GeV. Here, "Theory" refers to the points allowed by perturbativity, unitarity and the BFB conditions. And "Theory + Expt" denotes the points allowed by all the theoretical and experimental constraints. Right panel: Variation of $F_{Z,S}$ with tan $\beta$ for $M_{S_I} - M_{S_R} =$ 50 GeV, 100 GeV, 150 GeV in the $(M_{H^+}, M_{S_R}) = (200\,\text{GeV}, 800\,\text{GeV})$ case. The color coding is explained in the legends

$M_c = 1.6$ GeV, $M_\tau = 1.77$ GeV.[6] The array $M_{H^+} = 150$ GeV, 175 GeV, 200 GeV, 400 GeV, 600 GeV, 800 GeV and 1 TeV is chosen. Masses below 600 GeV however are omitted for Type-II in view of the flavour constraints. Variation of the absolute values of these form factors with respect to tan$\beta$ is depicted for Type-I and Type-II in Figs. 10 and 11. We have not shown the variation of $|F_\gamma|$ separately since it is not independent, and, is derivable from $|G_\gamma|$ through Eq. (3.4). Contributions to $H_Z$ and $H_\gamma$ come from the 2HDM fermionic part only. Expressions for the form factors can be found in Appendix A.5.

It can be inferred from Figs. 10 and 11 that the basic nature of the variations of $|F_Z|$, $|G_Z|$ and $|G_\gamma|$ with tan$\beta$ are the same for different $M_{H^+}$. While the variation of the color-octet contributions to these form factors with tan$\beta$ does not follow any particular trend,[7] the corresponding 2HDM contributions, for both Type-I and Type-II 2HDM, decrease with increasing tan $\beta$. This behaviour of the 2HDM is due to the $\lambda_{HH^+H^-} \propto \cot\beta$ property and the $\bar{t}bH^+$ interaction carrying the factor $\big(M_t \cot\beta + M_b \cot(\tan)\beta\big)/v$ in Type-I(II) 2HDM. A noticeable difference between the fermionic contributions of Type-I and II therefore is seen only when tan$\beta$ is on the higher side.

Figure 10 displays the variations of $|F_Z|$, $|G_Z|$ (include contributions from both 2HDM and color-octet), $|F_{Z,\text{2HDM}}|$, $|G_{Z,\text{2HDM}}|$ with tan $\beta$ for the values $M_{H^+} \geq 175$ GeV.[8] A similar variation is shown in Fig. 11, for $|G_{\gamma,\text{2HDM}}|$ and $|G_\gamma|$. For example, in case of $M_{H^+} = 200$ GeV, the 2HDM contribution to $|F_Z|$ measures $\simeq 0.018$ and this occurs for the lowest value of tan$\beta = 2$. The largest the colored sector can contribute for tan$\beta = 2$ is $\simeq 0.01$. The 2HDM contribution diminishes for higher tan$\beta$ and the colored scalars lead to higher relative enhancement. For tan$\beta = 8$, for instance, the maximal color-octet contribution ($\simeq 0.015$) is larger than the 2HDM contribution ($\simeq 0.0055$) by a factor $\sim 3$. More precisely, the enhancement in $F_Z$ from the color-octet increases from $\simeq 41\%$ to $\simeq 327\%$ when tan$\beta$ increases from 2 to 10. The relative enhancement in $|F_Z|$ coming from the colored scalars increases for a heavier charged Higgs. Raising $M_{H^+}$ implies in our case that $M_H$ and $M_A$ also increase accordingly. The overall 2HDM mass scale is raised and therefore the loop contribution starts diminishing. For $M_{H^+} = 600$ GeV in the Type-I 2HDM, while the maximum 2HDM contribution decreases from $\simeq 0.0066$ to $\simeq 0.0013$ as tan$\beta$ increases from 2 to 10, the color-octet contribution reaches $\simeq 0.02$. In other words, the relative enhancement from the colored scalars is $\simeq 230\%$ and $\simeq 950\%$ for tan$\beta = 2$ and 10 respectively. There is thus a ten-fold enhancement in this case. In case of Type-II for the same $M_{H^+}$, the 2HDM contribution is a bit higher owing to a different fermionic contribution and lies in the [0.002,0.008] range. The enhancement from the colored scalars is still substantial. The colored scalars provide a 117% (570%) relative enhancement for tan$\beta = 2$ (10).

[6] Here $M_t, M_b$ and $M_c$ are the running masses of $t, b, c$ at $\mu =$ 173.2 GeV ($t$-quark pole mass) in the $\overline{\text{MS}}$ scheme.

[7] The couplings $\lambda_{H^+S^-S_R}$ and $\lambda_{H^+S^-S_I}$ are not proportional to cot$\beta$ as read from Eqs. (5.5a) and (5.5b).

[8] $M_{H^+} = 150$ GeV was excluded while calculating the $H^+W^-Z$ form factors since it does not lead to on-shell $W^+$, $Z$.

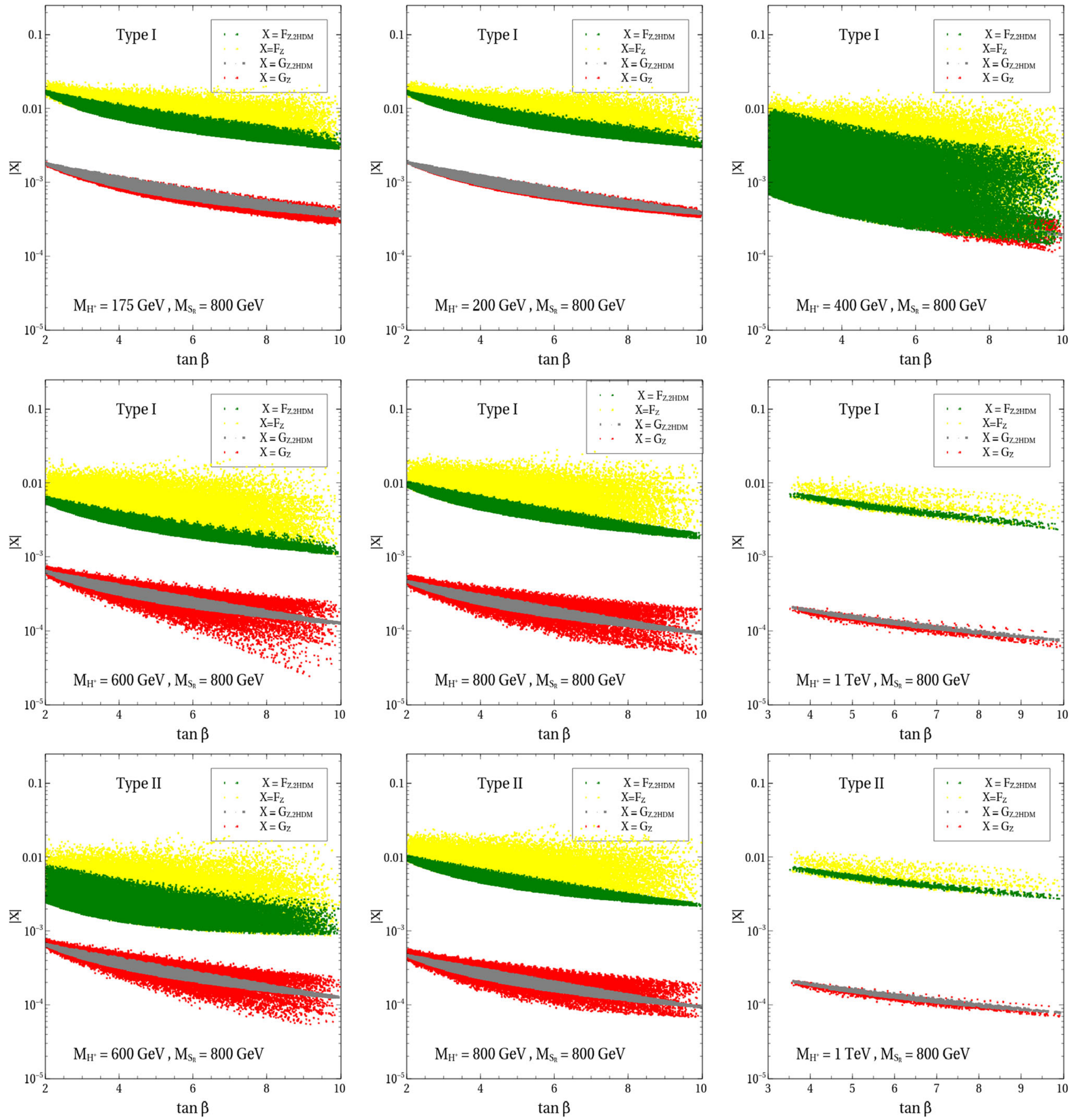

**Fig. 10** Variation of absolute value of the form factors $X = F_Z, G_Z, F_{Z,2\text{HDM}}, G_{Z,2\text{HDM}}$ with $\tan\beta$ for $M_{H^+} = 175, 200, 400, 600, 800, 1000$ GeV and $M_{S_R} = 800$ GeV for Type-I and Type-II 2HDM. Color coding is expressed in legends

On the other hand, such sizeable enhancements are not seen in the case of $|G_{Z(\gamma)}|$. The color-octet accounts for only $\simeq 10$ % of the total value of $G_\gamma$ for $M_{H^+} = 150$ GeV. For $M_{H^+} = 600$ GeV in case of Type-I, the color-octet contribution to $|G_{Z(\gamma)}|$ weighs up to $\simeq 1.3\times10^{-4}$ $(3.1\times10^{-4})$, while the maximum 2HDM contribution for a given $\tan\beta$ varies from $\simeq 1.4\times10^{-4}$ $(1.3\times10^{-4})$ to $\simeq 6.5\times10^{-4}$ $(6.4\times10^{-4})$. This translates to a 20% (50%) correction to $G_{Z\gamma}$ for $\tan\beta = 2$. The corresponding Type-II percentages are close as is revealed by Figs.10 and 11. Overall, $|F_Z|$ picks up a larger share of contribution from colored scalars than does $|G_{Z(\gamma)}|$ for all values of $M_{H^+}$.

We next come to a discussion of the $H^+ \to W^+ V$ branching fractions. The values of $M_{H^+}$ taken can be arranged into the piecewise intervals: (a) $M_{H^+} < M_W + M_Z$, (b) $M_W + M_Z \le M_{H^+} < M_t + M_b$, and, (c) $M_t + M_b \le M_{H^+} < 1$ TeV. Intervals (a) and (b) apply exclusively to Type-I 2HDM while both Type-I and Type-II 2HDM can be accommodated in (c). The choice of intervals (a) and (b) are intended to maximise the branching fraction for $H^+ \to W^+V$ by kinematically blocking certain tree level decays. Interval (a) blocks $H^+ \to W^+Z,\ t\bar{b}$, while (b), a rather narrow and fine-tuned mass range, forbids $t\bar{b}$.

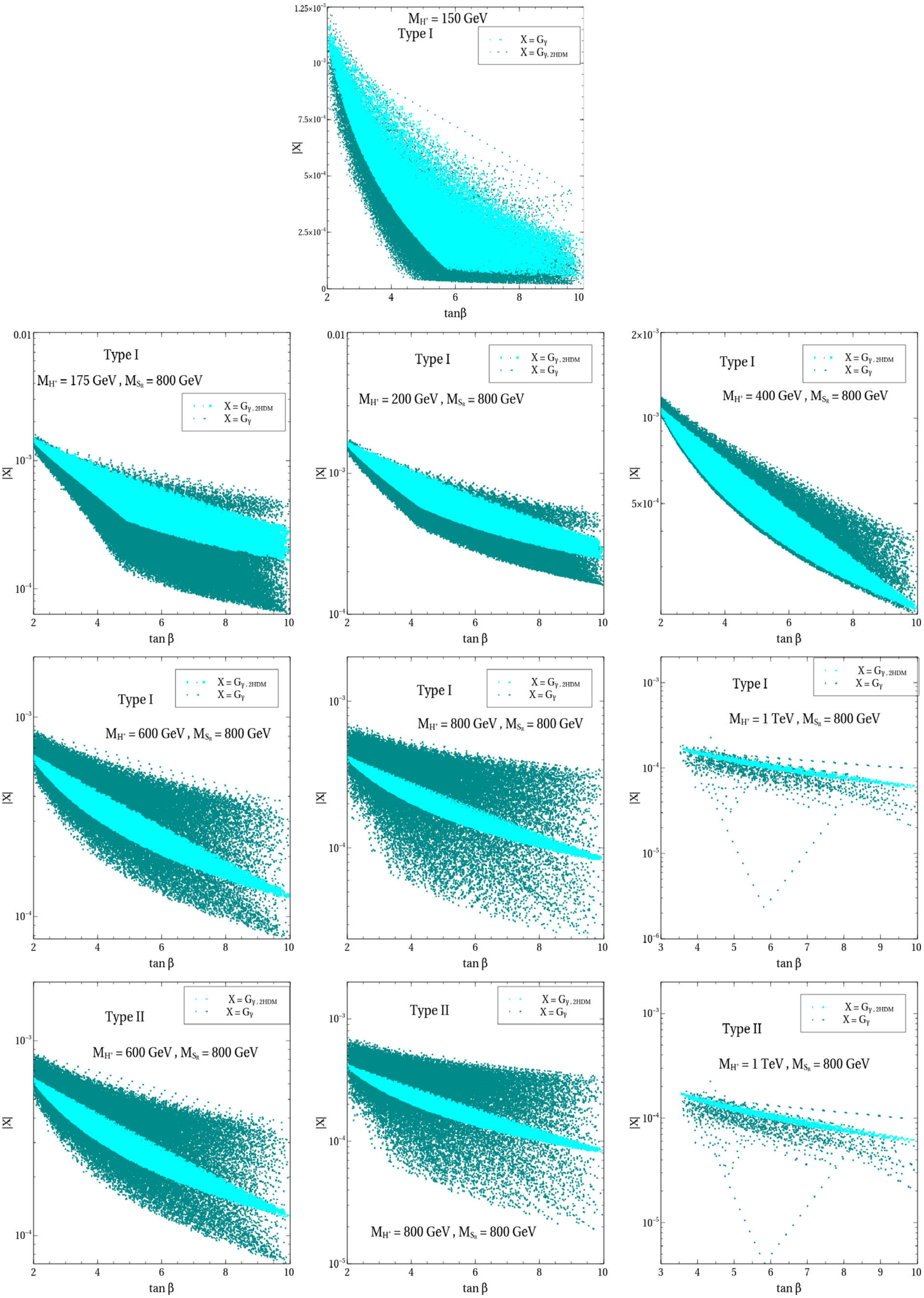

**Fig. 11** Variation of absolute value of the form factors $X = G_\gamma, G_{\gamma,2\mathrm{HDM}}$ with $\tan\beta$ for $M_{H^+} = 175, 200, 400, 600, 800, 1000$ GeV and $M_{S_R} = 800$ GeV for Type-I and Type-II 2HDM. Color coding is expressed in legends

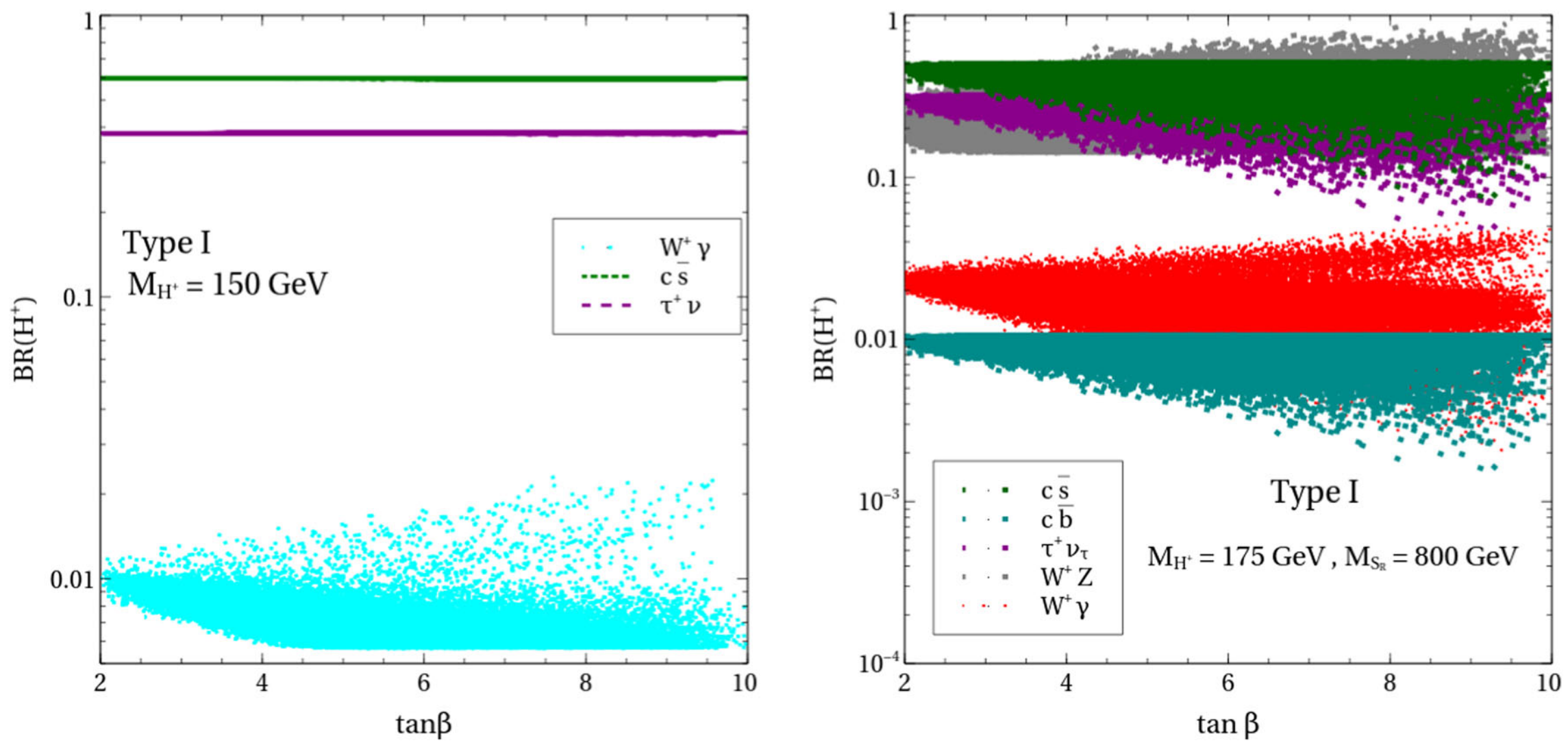

**Fig. 12** Variation of the branching ratios of $H^+$ with $\tan\beta$ for $M_{H^+} = 150$, 175 GeV for Type-I 2HDM

Figure 12 generates idea on how large the radiative $H^+ \to W^+V$ decay fractions can be in intervals (a) and (b). With $H^+ \to c\bar{s}, \tau^+\nu_\tau$ constituting the dominant share, the maximum $H^+ \to W^+\gamma$ branching ratio an $H^+$ of mass 150 GeV can attain is on the upward of 2% when $\tan\beta \gtrsim 7$. Whereas for $M_{H^+} = 175$ GeV, $H^+ \to W^+Z$ can compete with the $c\bar{s}, \tau^+\nu_\tau$ modes and even become the dominant one for $\tan\beta$ exceeding 4. In fact, it gets as high as $\simeq 85\%$ for $\tan\beta = 8$.[9] The $W^+\gamma$ branching fraction gets suppressed whenever the $W^+Z$ channel opens up. The former is found to be at best $\simeq 5\%$ for $M_{H^+} = 175$ GeV.

The $t\bar{b}$ mode opens up in interval (c) (Fig. 13 shows the branching ratios.) and the corresponding branching ratio becomes the dominant one ($\sim 99\%$). Nonetheless, $H^+ \to W^+Z$ always surpasses $H^+ \to c\bar{s}, \tau^+\nu_\tau$ for Type-I and can be $\mathcal{O}(1\%)$ for an heavy $H^+$. We find that it can be up to $\simeq 3\%$ for $M_{H^+} = 800$ GeV and 1 TeV. The $c\bar{s}, \tau^+\nu_\tau$ modes however become competing for Type-II with even overtaking $W^+Z$. This difference with Type-I is solely traced to the different structure of the Yukawa couplings in the two cases. For Type-II also, the $H^+ \to W^+Z$ branching fraction remains in the same ballpark as in Type-I. We add here that the $H^+ \to W^+\gamma$ branching ratio is negligibly small and hovers in the range $[10^{-6}, 10^{-4}]$ for both Type-I and Type-II for all $M_{H^+}$ in interval (c). One must note that while $H^+ \to W^+H,\ W^+A$ are kinematically closed upon setting $M_{H^+} = M_H$ and $M_A > M_{H^+}$ in the scan, $H^+ \to W^+h$ vanishes for the $c_{\beta-\alpha} = 0$ limit chosen.

Lastly, in Fig. 14, we show the impact of raising the color-octet mass scale on the form factors. The form factors diminish in size and approach the limiting value given by Eqs. (5.3a) and (5.3b). The branching fraction diminishes accordingly upon increasing $M_{S_R}$.

## 6 Cross sections at $pp$ colliders

In this section, we examine the prospects of probing the $H^+W^-Z$ interaction at $pp$ collisions. We identify the following three processes that feature the $H^+W^-Z$ coupling at the production vertex, the decay vertex, or both: (a) $pp \to H^+jj \to W^+Zjj$,[10] (b) $pp \to \bar{t}H^+ \to \bar{t}W^+Z$, (c) $pp \to H^+jj \to t\bar{b}jj$ (see Fig. 15 for the Feynman diagrams.). The corresponding cross sections are computed for the following sample points (Table 2).

We use `MadGraph5 aMC@NLO` [92] with the `NN23LO1` parton distribution function (PDF) to compute the LHC cross sections for these processes.[11] The corresponding numbers for $\sqrt{s} = 14$ TeV and 100 TeV [93–96] are respectively given in Tables 3 and 4. The $pp \to H^+jj$ process is generated using the 5-flavour scheme while imposing the cuts $p_T^j > 20$ GeV, $|\eta_j| < 0.5$ and $\Delta R_{jj} > 0.4$.

[9] The very tiny $H^+ \to W^+f\bar{f}$ 3-body decay has been neglected in the calculation.

[10] Experimental analyses of this process are available in [90,91].

[11] An LO PDF such as `NN23LO1` must be used for computing LO cross sections like what we have here.

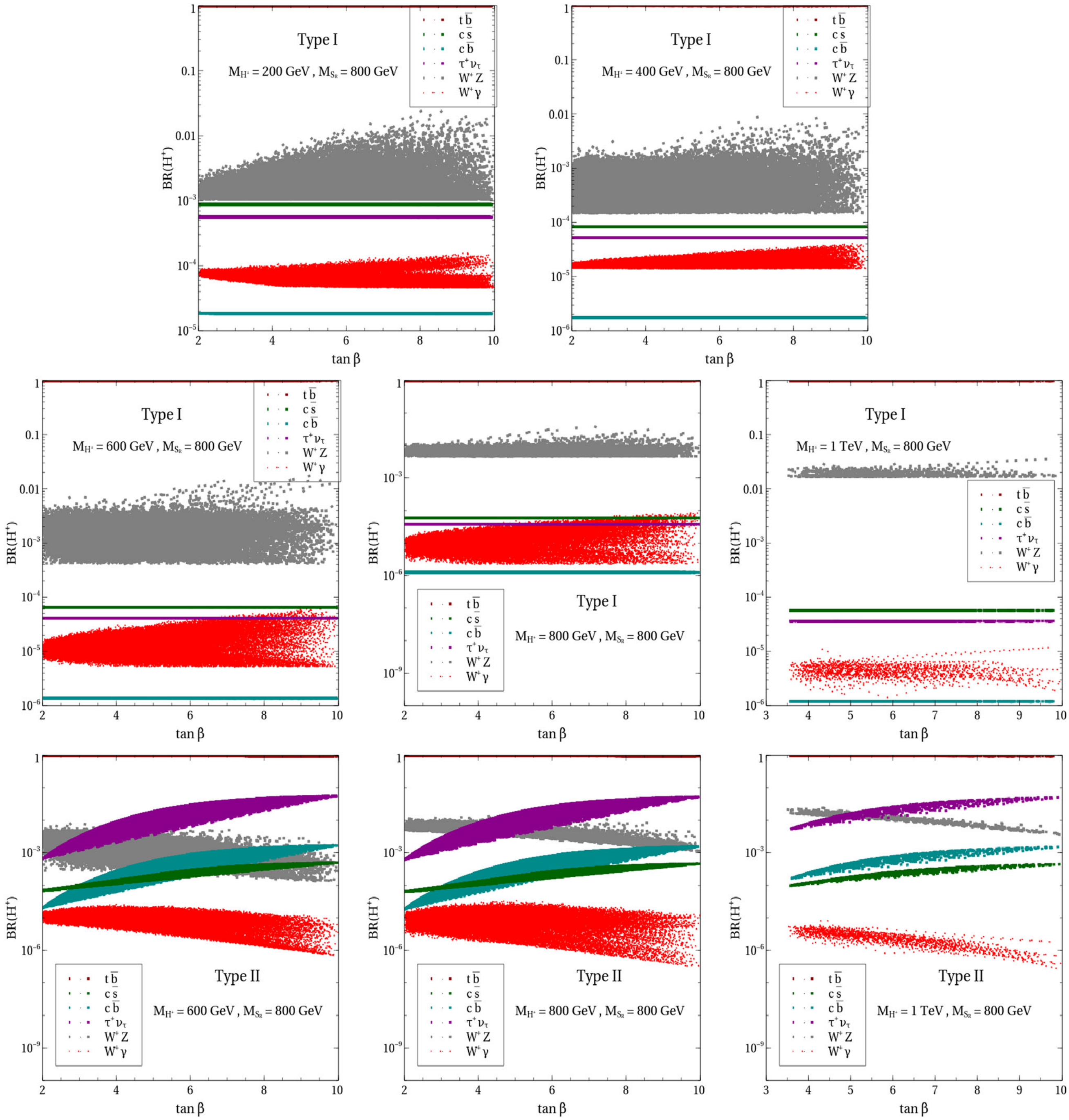

**Fig. 13** Variation of branching ratio of $H^+$ into different decay modes with $\tan\beta$ for $M_{H^+} = 200, 400, 600, 800, 1000$ GeV and $M_{S_R} = 800$ GeV for Type-I and Type-II 2HDM

**Table 2** Sample parameter points proposed to compute cross sections at $pp$ colliders

| Sample point | $M_{H^+}$ (GeV) | $\tan\beta$ | $\lvert F_Z \rvert$ | $\lvert G_Z \rvert$ | BR $(H^+ \to W^+ Z)$ | BR $(H^+ \to t\bar{b})$ |
|---|---|---|---|---|---|---|
| SP1 | 175 | 8.12 | $2.07 \times 10^{-2}$ | $3.93 \times 10^{-4}$ | $8.54 \times 10^{-1}$ | 0 |
| SP2 | 200 | 6.05 | $1.88 \times 10^{-2}$ | $5.88 \times 10^{-4}$ | $1.27 \times 10^{-2}$ | $9.98 \times 10^{-1}$ |
| SP3 | 600 | 5.65 | $1.85 \times 10^{-2}$ | $9.58 \times 10^{-5}$ | $5.57 \times 10^{-3}$ | $9.99 \times 10^{-1}$ |

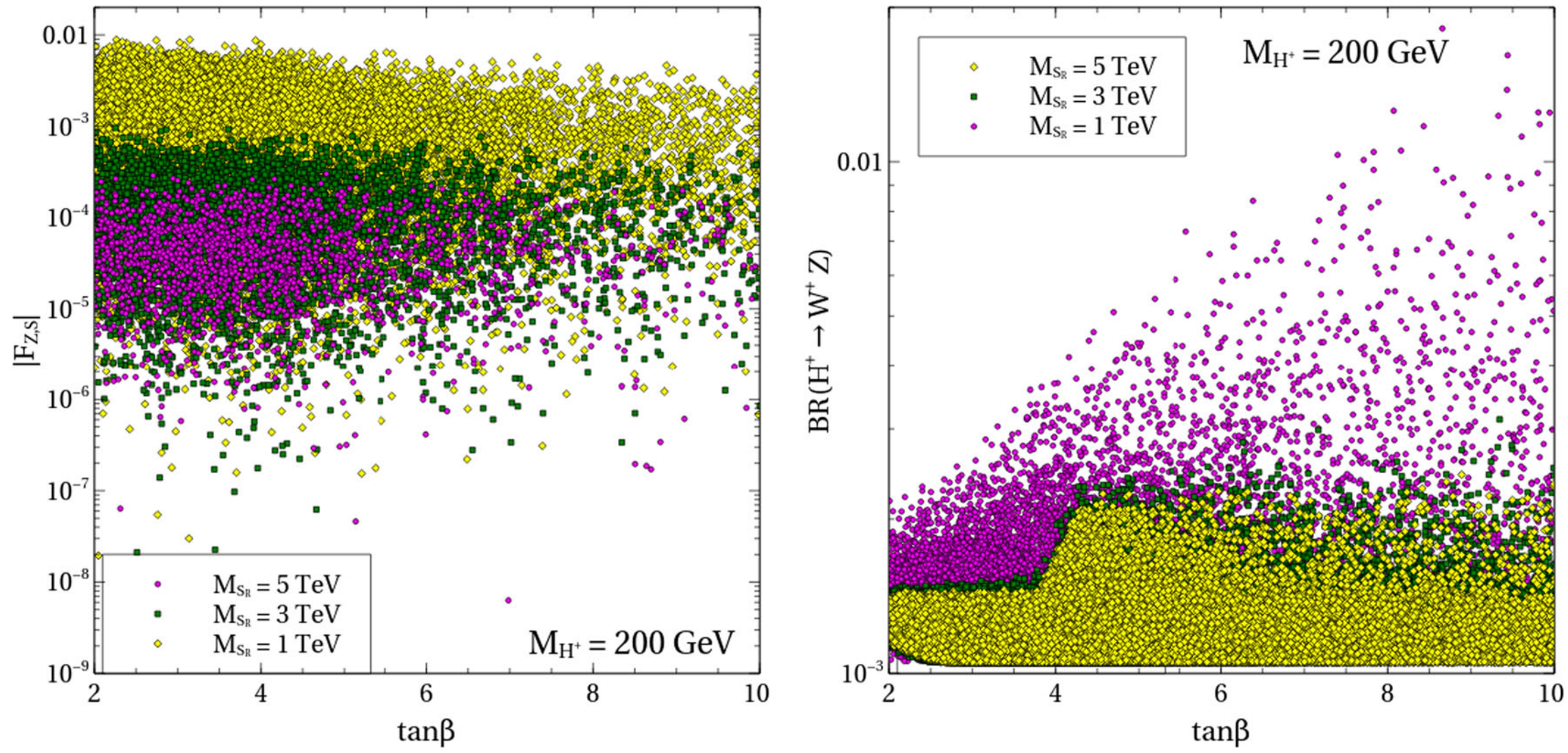

**Fig. 14** $|F_{Z,S}|$ (left) and BR($H^+ \to W^+Z$) (right) for $M_{H^+} = 200$ GeV and $M_{S_R} = 1$ TeV, 3 TeV and 5 TeV

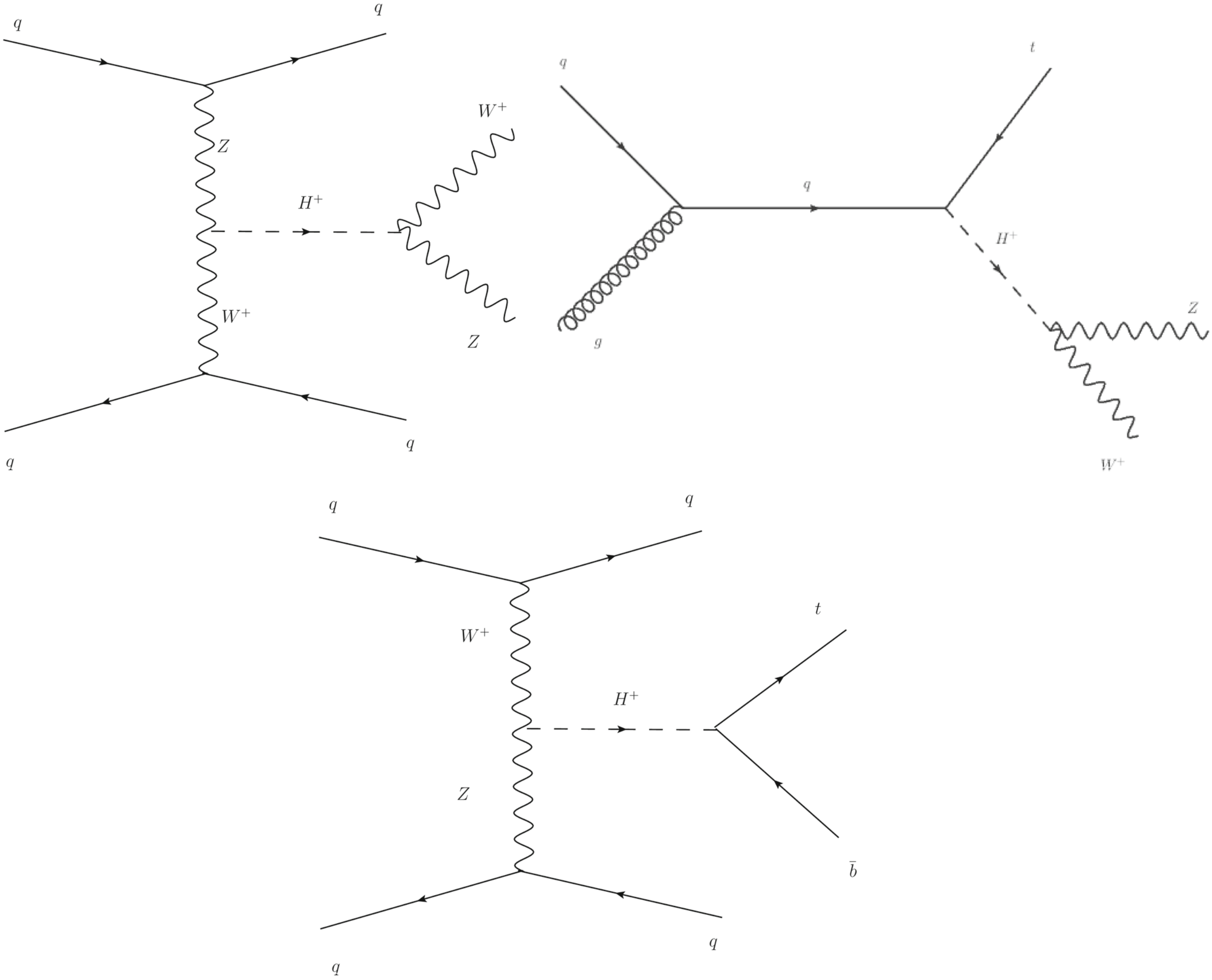

**Fig. 15** Feynman diagrams for the processes (**a**) (top left), (**b**) (top right), (**c**) (bottom)

**Table 3** Cross sections of processes (a), (b), (c) at 14 TeV LHC

| Sample point | $\sigma_{pp\to H^+jj\to W^+Zjj}$ (fb) | $\sigma_{pp\to\bar{t}H^+\to tW^+Z}$(fb) | $\sigma_{pp\to H^+jj\to t\bar{b}jj}$ (fb) |
|---|---|---|---|
| SP1 | 0.498 | 309.71 | 0 |
| SP2 | $8.23\times10^{-4}$ | 0.51 | $6.46\times10^{-2}$ |
| SP3 | $3.33\times10^{-4}$ | 0.255 | $5.97\times10^{-2}$ |

**Table 4** Cross sections of processes (a), (b), (c) at 100 TeV FCC-hh

| Sample point | $\sigma_{pp\to H^+jj\to W^+Zjj}$ (fb) | $\sigma_{pp\to\bar{t}H^+\to tW^+Z}$ (fb) | $\sigma_{pp\to H^+jj\to t\bar{b}jj}$ (fb) |
|---|---|---|---|
| SP1 | 5.254 | $1.421\times10^{4}$ | 0 |
| SP2 | 0.015 | 42.075 | 1.179 |
| SP3 | $5.687\times10^{-3}$ | 21.115 | 1.019 |

The most hefty event-yield is obviously obtained from $pp \to \bar{t}H^+$ since this is a purely tree level production process in contrast to the others that involve the one-loop $H^+W^-Z$ vertex. In SP3 for instance, $H^+$ having even a sizeable mass of 600 GeV, produced in association with a $\bar{t}$, can decay to a gauge boson pair $6\times10^5$ times in $pp$ collisions at $\sqrt{s}$ = 100 TeV with an integrated luminosity $L = 30$ ab$^{-1}$. The numbers are definitely higher for a lighter $H^+$ in SP1 and SP2. A complete treatment however would entail considering the decays of the gauge bosons, and, a careful analysis of the signal and the SM backgrounds.

Despite offering lower event-yields than process (c), processes (a) and (d) are interesting from a kinematical perspective. Since all three involve the $H^+W^-Z$ interaction at the production vertex, the momentum dependent term in the amplitude (proportional to $G_Z$) becomes potentially important especially when a large momentum transfer is involved. This momentum dependence will distort various kinematical distributions to a degree controlled by the value of $G_Z$. Further amongst these, $H^+$ is produced via vector boson fusion (VBF) in (a) and (c) followed by decay to $t\bar{b}$ and $W^+Z$ respectively. It is then understood that (c) is useful to probe the $M_{H^+} > M_t + M_b$ region. As an example, SP3 would lead to the production of $3\times10^4$ $t\bar{b}$ pairs at the 100 TeV FCC-hh. On the other hand, the $M_W + M_Z < M_{H^+} < M_t + M_b$ window can be probed through (a). In fact, its cross section for SP1 is very similar to that of (c) for SP2. Further, when both $W^+$ and $Z$ decay leptonically, process (c) leads to less hadronic activity compared to (a) enhancing the discernibility of the latter at the LHC.

For $\tan\beta = 8$, $\sigma_{pp\to\bar{t}H^+\to\bar{t}W^+\gamma}$ measures up to 29 fb ( 1017 fb) at 14 TeV (100 TeV) $pp$-collisions. One must note that the aforementioned cross sections are much smaller than what would have been in case of the Georgi-Machacek (GM) model [97–109], or, scalar-triplet extensions of the MSSM [110–112] that predict $F_Z \sim \mathcal{O}(0.1)$ at the tree level itself.[12] However, Machine Learning techniques can be invoked to perform state-of-the-art event analyses and possibly distinguish the present scenario from the GM model by probing the momentum dependence of production amplitude discussed above.

## 7 Summary and conclusions

The weak isospin symmetry of the kinetic terms in models with $SU(2)_L$ scalar doublets forbids the $H^+W^-Z(\gamma)$ vertex at the tree level. However, radiative effects can give rise to this vertex at the one-loop level. It is therefore imperative to estimate the size of this vertex in such models and probe its observability at the energy frontier.

In this work, we have quantified the strength of the $H^+W^-Z(\gamma)$ interaction in context of an extended scalar sector comprising two color-singlet $SU(2)_L$ doublets and a color-octet $SU(2)_L$ doublet. While a Yukawa interaction between the SM fermions and the color-octet $SU(2)_L$ doublet is not relevant for the study, the corresponding ones involving the color singlet doublets are taken similar to the flavour conserving Type-I and Type-II 2HDMs. The analysis takes into account all possible constraints. On the theoretical side, the constraints come from perturbativity of quartic couplings, unitarity of the Lee-Quigg-Thacker (LQT) eigenvalues and bounded-from-below criteria of the scalar potential. Experimental constraints include the bound on the charged Higgs mass $M_{H^+}$ from $B \to X_s\gamma$, Higgs signal strengths, oblique parameters and exclusion limits from LHC. We adopted the non-linear gauge to gauge-out the unphysical $G^+W^-Z(\gamma)$ vertex. This simplified our calculation considerably. In order to throw more insight into the analysis, we presented simpli-fied expressions of the form factors that are demonstrative of the decoupling/non-decoupling behaviour. We summarise below our key observations from the formulae.

[12] This is true also in some custodial variants of the GM model detailed in [105].

- The one-loop form factors borne out of the color-octet carry a color factor of 8.
- Amongst the various color-octet form factors, $F_{Z,S}$ clearly displays a non-decoupling behaviour. That is, it does not vanish in the $M_{S_R} \longrightarrow \infty$ limit. This is therefore identified as the source of a non-decoupled loop contribution to $H^+ \to W^+Z$ partial width. On the contrary, $F_{\gamma,S}$ and $G_{\gamma,S}$ decouple thereby inducing a similar behaviour in $H^+ \to W^+\gamma$ width.
- The coupling $\lambda_{H^+S^-S_I}$ is sensitive to $(M_{S_I}^2 - M_{S_R}^2)$ and so are the color-octet form factors.

We have evaluated the various form factors and the $H^+ \to W^+Z(\gamma)$ branching ratio for $M_{H^+} \in [150\text{ GeV}, 1\text{ TeV}]$. The color-octet is found to enhance $F_Z$ by up to a factor of $\sim 3$ with respect to the purely 2HDM value. While the 2HDM contribution registers a decrease with increasing tan$\beta$, the color-octet contribution does not have such a behaviour. A larger relative enhancement is thus seen for higher tan$\beta$. The $H^+ \to W^+Z$ branching fraction is found to touch the $\mathcal{O}(1)$ % ball-park for $M_{H^+} > 200$ GeV, the maximum being $\simeq 3\%$ for $M_{H^+} = 1$ TeV. On the other hand, the same branching fraction is seen to be as high as $\sim 90$ % for $M_{H^+} = 175$ GeV in which case the $H^+ \to t\bar{b}$ mode is not open. We have also discussed the sensitivity of the $H^+W^-V$ vertex at hadron colliders in the end. The $pp \to \bar{t}H^+ \to \bar{t}W^+V$ is found to offer a handsome event-yield. However, processes that involve the $H^+W^-Z$ vertex at the production level can also be interesting from a kinematic viewpoint.

A closing remark is in order. Since the dominant contribution to the amplitude comes from the virtual effect of colored scalars, QCD corrections could be important. This point warrants further study and will be presented in a future work.

**Acknowledgements** NC is financially supported by NCTS (National Center for Theoretical Sciences), and, IISc (Indian Institute of Science) through the C.V.Raman postdoctoral fellowship. IC acknowledges support from DST, India, under grant number IFA18-PH214 (INSPIRE Faculty Award), and, hospitality extended by NCTS and IISc while this work was in progress. Both NC and IC are thankful to Indian Institute of Technology Guwahati for organising WHEPP (Workshop on High Energy Physics Phenomenology) XVI where a part of the work was completed.

**Data Availability Statement** This manuscript has no associated data or the data will not be deposited. [Authors' comment: The data we used in our manuscript to compare with our model estimations are already public from the ATLAS and CMS experiments.]

Funded by SCOAP[3].

## A Appendix

This section contains various analytical expressions related to form factors and partial widths.

### A.1 Decay widths of $h$

#### A.1.1 $h \longrightarrow \gamma\gamma$

$$\mathcal{M}^{\text{2HDM+S}}_{h\to\gamma\gamma} = \sum_f N_f Q_f^2 f_{hff} A_{1/2}\left(\frac{M_\phi^2}{4M_f^2}\right) + f_{\phi VV} A_1\left(\frac{M_h^2}{4M_W^2}\right) + \frac{\lambda_{hH^+H^-}v}{2M_{H^+}^2} A_0\left(\frac{M_h^2}{4M_{H^+}^2}\right) + N_S \frac{\lambda_{hS^+S^-}v}{2M_{S^+}^2} A_0\left(\frac{M_\phi^2}{4M_{S^+}^2}\right), \tag{A.1a}$$

$$\Gamma^{\text{2HDM+S}}_{h\to\gamma\gamma} = \frac{G_F\alpha^2 M_h^3}{128\sqrt{2}\pi^3}\left|\mathcal{M}^{\text{2HDM+S}}_{h\to\gamma\gamma}\right|^2. \tag{A.1b}$$

where $G_F$ and $\alpha$ denote respectively the Fermi constant and the QED fine-structure constant. The loop functions are listed below.

$$A_{1/2}(x) = \frac{2}{x^2}\big((x + (x-1)f(x)\big), \tag{A.2a}$$

$$A_1(x) = -\frac{1}{x^2}\big((2x^2 + 3x + 3(2x-1)f(x)\big), \tag{A.2b}$$

$$A_0(x) = -\frac{1}{x^2}\big(x - f(x)\big), \tag{A.2c}$$

$$\text{with } f(x) = \arcsin^2(\sqrt{x}); \quad x \le 1$$
$$= -\frac{1}{4}\left[\log\frac{1+\sqrt{1-x^{-1}}}{1-\sqrt{1-x^{-1}}} - i\pi\right]^2; \quad x > 1. \tag{A.2d}$$

where $A_{1/2}(x)$, $A_1(x)$ and $A_0(x)$ are the respective amplitudes for the spin-$\frac{1}{2}$, spin-1 and spin-0 particles in the loop.

$$f_{htt} = \frac{\cos\alpha}{\sin\beta},\ f_{hVV} = \sin(\beta-\alpha). \tag{A.3a}$$

*A.1.2 $h \longrightarrow gg$*

$$\begin{aligned}\mathcal{M}^{\text{2HDM+S}}_{h\to gg} &= \sum_f \frac{3}{4} f_{\phi ff} A_{1/2}\left(\frac{M_h^2}{4M_f^2}\right) \\ &+ \frac{9\lambda_{hS_RS_R} v}{8M_{S_R}^2} A_0\left(\frac{M_h^2}{4M_{S_R}^2}\right) \\ &+ \frac{9\lambda_{hS_IS_I} v}{8M_{S_I}^2} A_0\left(\frac{M_h^2}{4M_{S_I}^2}\right) \\ &+ \frac{9\lambda_{hS_RS_R} v}{4M_{S^+}^2} A_0\left(\frac{M_h^2}{4M_{S^+}^2}\right), \end{aligned} \tag{A.4a}$$

$$\Gamma^{\text{2HDM+S}}_{h\to gg} = \frac{G_F\alpha_s^2 M_h^3}{36\sqrt{2}\pi^3}\left|\mathcal{M}^{\text{2HDM+S}}_{h\to gg}\right|^2. \tag{A.4b}$$

Here, $\alpha_s$ refers to the strong coupling constant.

*A.1.3 $h \longrightarrow Z\gamma$*

$$\begin{aligned}\mathcal{M}^{\text{2HDM+S}}_{h\to Z\gamma} &= \sum_f N_f Q_f \frac{\left(2I_3^f - 4Q_f s_W^2\right)}{c_W} f_{hff} B_{1/2} \\ &\times \left(\frac{M_h^2}{4M_t^2}, \frac{M_Z^2}{4M_t^2}\right) + f_{hVV} B_1\left(\frac{M_h^2}{4M_W^2}, \frac{M_Z^2}{4M_W^2}\right) \\ &+ \frac{\lambda_{hH^+H^-} v}{2M_{H^+}^2} B_0\left(\frac{M_h^2}{4M_{H^+}^2}, \frac{M_Z^2}{4M_{H^+}^2}\right) \\ &+ N_S \frac{\lambda_{hS^+S^-} v}{2M_{S^+}^2} B_0\left(\frac{M_h^2}{4M_{S^+}^2}, \frac{M_Z^2}{4M_{S^+}^2}\right), \end{aligned} \tag{A.5a}$$

$$\Gamma^{\text{2HDM+S}}_{h\to Z\gamma} = \frac{G_F^2\alpha M_W^2 M_h^3}{64\pi^4}\left(1 - \frac{M_Z^2}{M_h^2}\right)^3 \left|\mathcal{M}^{\text{2HDM+S}}_{h\to Z\gamma}\right|^2, \tag{A.5b}$$

$$I_1(x,y) = -\frac{1}{2(x-y)} + \frac{f(x)-f(y)}{2(x-y)^2} + \frac{y\{g(x)-g(y)\}}{(x-y)^2}, \tag{A.6a}$$

$$I_2(x,y) = \frac{f(x)-f(y)}{2(x-y)}, \tag{A.6b}$$

$$B_0(x,y) = I_1(x,y), \tag{A.6c}$$

$$B_{1/2}(x,y) = I_1(x,y) - I_2(x,y), \tag{A.6d}$$

$$B_1(x,y) = c_W\left\{4\left(3 - \frac{s_W^2}{c_W^2}\right) I_2(x,y) + \left((1+2x)\frac{s_W^2}{c_W^2} - (5+2x)\right) I_1(x,y)\right\}, \tag{A.6e}$$

$$\begin{aligned} g(x) &= \sqrt{x^{-1}-1}\,\arcsin(x); \quad x \le 1 \\ &= \frac{\sqrt{1-x^{-1}}}{2}\left[\log\frac{1+\sqrt{1-x^{-1}}}{1-\sqrt{1-x^{-1}}} - i\pi\right]; \quad x > 1 \end{aligned} \tag{A.6f}$$

## A.2 $H^+$ decay widths

- $H^+ \to t\bar{b}$

$$\begin{aligned}\mathcal{M}_{H^+\to t\bar{b}} &= \frac{N_t}{v^2}\left[A_{tb}^2\{M_{H^+}^2 - (M_t+M_b)^2\} + B_{tb}^2\{M_{H^+}^2 - (M_t-M_b)^2\}\right], \\ \Gamma_{H^+\to t\bar{b}} &= \frac{|\mathcal{M}|^2}{16\pi M_{H^+}}\sqrt{\lambda\left(1, \frac{M_t^2}{M_{H^+}^2}, \frac{M_b^2}{M_{H^+}^2}\right)}, \\ &\quad \text{for} \quad M_{H^+} > M_t + M_b, \\ &= 0, \quad \text{for} \quad M_{H^+} < M_t + M_b. \end{aligned} \tag{A.7}$$

Here,

$$\lambda(x,y,z) = x^2 + y^2 + z^2 - 2xy - 2yz - 2zx, \tag{A.8}$$

$$A_{tb} = M_b\zeta_b + M_t\zeta_t, \tag{A.9}$$

$$B_{tb} = M_b\zeta_b - M_t\zeta_t. \tag{A.10}$$

Here, $\zeta_t$ and $\zeta_b$ are the scale factors with respect to SM. For type-I (type-II) 2HDM $\zeta_t = \cot\beta\ (\cot\beta), \zeta_b = -\cot\beta\ (\tan\beta)$.

- $H^+ \to c\bar{s}$
  Decay width formula is same as $H^+ \to t\bar{b}$, with the replacement $M_t \to M_c,\ M_b \to M_s$.
- $H^+ \to \tau^+\nu_\tau$
  Decay width formula is same as $H^+ \to t\bar{b}$, with the replacement $M_t \to M_\tau,\ M_b \to 0$ , since $M_{\nu_\tau} = 0$.
- $H^+ \to W^+\phi,\ \phi = h, H, A$

$$\Gamma_{H^+\to W^+\phi} = \frac{a^2}{16\pi v^2 M_{H^+}^3}\lambda \times \left(1, \frac{M_W^2}{M_{H^+}^2}, \frac{M_\phi^2}{M_{H^+}^2}\right)^{3/2} \tag{A.11}$$

where $a = c_{\beta-\alpha}$ for $\phi = h$ and $a = s_{\beta-\alpha}$ for $\phi = H$

## A.3 Scalar trilinear vertices

$$\lambda_{hS^+S^-} = \frac{v}{2}(-\nu_1 c_\beta s_\alpha + \omega_1 s_\beta c_\alpha + \kappa_1 c_{\beta+\alpha}), \tag{A.12a}$$

$$\begin{aligned}\lambda_{hS_RS_R} &= \frac{v}{2}\{-(\nu_1+\nu_2+2\nu_3)c_\beta s_\alpha \\ &+ (\omega_1+\omega_2+2\omega_3)s_\beta c_\alpha\end{aligned}$$

$$+(\kappa_1+\kappa_2+\kappa_3)c_{\beta+\alpha}\}, \quad \text{(A.12b)}$$

$$\lambda_{hS_IS_I} = \frac{v}{2}\{-(\nu_1+\nu_2-2\nu_3)c_\beta s_\alpha + (\omega_1+\omega_2-2\omega_3)s_\beta c_\alpha + (\kappa_1+\kappa_2-\kappa_3)c_{\beta+\alpha}\}, \quad \text{(A.12c)}$$

$$\lambda_{H^+S^-S_R} = \frac{1}{4}v\{\sin\beta\ \cos\beta(-\nu_2-2\nu_3+\omega_2+2\omega_3) + (\kappa_2+\kappa_3)\cos 2\beta\}, \quad \text{(A.12d)}$$

$$\lambda_{H^+S^-S_I} = \frac{1}{4}v\{\sin\beta\ \cos\beta(-\nu_2+2\nu_3+\omega_2-2\omega_3) + (\kappa_2-\kappa_3)\cos 2\beta\}, \quad \text{(A.12e)}$$

$$\lambda_{hH^+H^-} = v\big\{\big(-\lambda_3 c_\beta^3+(-\lambda_1+\lambda_4+\lambda_5)s_\beta^2 c_\beta\big)s_\alpha + \big(\lambda_3 s_\beta^3+(\lambda_2-\lambda_4-\lambda_5)c_\beta^2 s_\beta\big)c_\alpha\big\}, \quad \text{(A.12f)}$$

$$\lambda_{HH^+H^-} = \cos\alpha\{v\sin^2\beta\cos\ \beta(\lambda_1-\lambda_4-\lambda_5)+\lambda_3 v\cos^3\beta\} + \sin\alpha\ \sin\beta\{v\cos^2\beta(\lambda_2-\lambda_4-\lambda_5)+\lambda_3 v\sin^2\beta\} \quad \text{(A.12g)}$$

$$\lambda_{hH^+G^-} = \frac{v}{4}\{(\lambda_2-\lambda_3+\lambda_4+\lambda_5)c_\beta c_\alpha + (-\lambda_2+\lambda_3+\lambda_4+\lambda_5)c_{3\beta}c_\alpha + (\lambda_1-\lambda_3+\lambda_4+\lambda_5)s_\beta s_\alpha - (-\lambda_1+\lambda_3+\lambda_4+\lambda_5)s_{3\beta}s_\alpha\} \quad \text{(A.12h)}$$

$$\lambda_{HH^+G^-} = \frac{v}{4}\{(\lambda_2-\lambda_3+\lambda_4+\lambda_5)c_\beta s_\alpha + (-\lambda_2+\lambda_3+\lambda_4+\lambda_5)c_{3\beta}s_\alpha - (\lambda_1-\lambda_3+\lambda_4+\lambda_5)s_\beta c_\alpha + (-\lambda_1+\lambda_3+\lambda_4+\lambda_5)s_{3\beta}c_\alpha\} \quad \text{(A.12i)}$$

### A.4 Passarino–Veltman functions

Throughout the analysis the loop functions have been expressed in terms of Passarino–Veltman functions [113]. The integral formulae of the functions relevant for our calculations, are illustrated below:

$$B_0(p^2;M_1,M_2) = \int\frac{d^dk}{i\pi^2}\frac{1}{(k^2-M_1^2)((k+p)^2-M_2^2)}, \quad \text{(A.13a)}$$

$$p^\mu B_1(p^2;M_1,M_2) = \int\frac{d^dk}{i\pi^2}\frac{k^\mu}{(k^2-M_1^2)((k+p)^2-M_2^2)}. \quad \text{(A.13b)}$$

We also have

$$C_0(p_1^2,p_2^2,q^2;M_1,M_2,M_3) = \int\frac{d^dk}{i\pi^2}\frac{1}{(k^2-M_1^2)((k+p_1)^2-M_2^2)(k+q)^2-M_3^2)} \quad \text{(A.14a)}$$

$$(p_1^\mu C_{11}+p_2^\mu C_{12})(p_1^2,p_2^2,q^2;M_1,M_2,M_3) = \int\frac{d^dk}{i\pi^2}\frac{k^\mu}{(k^2-M_1^2)((k+p_1)^2-M_2^2)(k+q)^2-M_3^2)} \quad \text{(A.14b)}$$

$$(p_1^\mu p_1^\nu C_{21}+p_2^\mu p_2^\nu C_{22}+p_1^\mu p_2^\nu C_{23}+g^{\mu\nu}C_{24})(p_1^2,p_2^2,q^2;M_1,M_2,M_3) = \int\frac{d^dk}{i\pi^2}\frac{k^\mu k^\nu}{(k^2-M_1^2)((k+p_1)^2-M_2^2)(k+q)^2-M_3^2)} \quad \text{(A.14c)}$$

The Passarino–Veltman functions have the closed forms below.

$$B_0(p^2;M_1,M_2) = \text{div} - \int_0^1 dx\,\ln\Delta_B, \quad \text{(A.15a)}$$

$$B_1(p^2;M_1,M_2) = -\frac{\text{div}}{2} + \int_0^1 dx(1-x)\ln\Delta_B, \quad \text{(A.15b)}$$

$$\times C_0(p_1^2,p_2^2,q^2;M_1,M_2,M_3) = -\int_0^1 dx\int_0^1 dy\frac{y}{\Delta_C}, \quad \text{(A.15c)}$$

$$\times C_{11}(p_1^2,p_2^2,q^2;M_1,M_2,M_3) = -\int_0^1 dx\int_0^1 dy\frac{y(xy-1)}{\Delta_C}, \quad \text{(A.15d)}$$

$$\times C_{12}(p_1^2,p_2^2,q^2;M_1,M_2,M_3) = -\int_0^1 dx\int_0^1 dy\frac{y(y-1)}{\Delta_C}, \quad \text{(A.15e)}$$

$$\times C_{21}(p_1^2,p_2^2,q^2;M_1,M_2,M_3) = -\int_0^1 dx\int_0^1 dy\frac{y(1-xy)^2}{\Delta_C}, \quad \text{(A.15f)}$$

$$\times C_{22}(p_1^2,p_2^2,q^2;M_1,M_2,M_3) = -\int_0^1 dx\int_0^1 dy\frac{y(1-y)^2}{\Delta_C}, \quad \text{(A.15g)}$$

$$\times C_{23}(p_1^2,p_2^2,q^2;M_1,M_2,M_3) = -\int_0^1 dx\int_0^1 dy\frac{y(1-xy)(1-y)}{\Delta_C}, \quad \text{(A.15h)}$$

$$\times C_{24}(p_1^2,p_2^2,q^2;M_1,M_2,M_3) = \frac{\text{div}}{4} - \frac{1}{2}\int_0^1 dx\int_0^1 dy\ y\ln\Delta_C. \quad \text{(A.15i)}$$

Where,

$$\Delta_B = -x(1-x)p^2 + xM_1^2 + (1-x)M_2^2, \quad \text{(A.16a)}$$

$$\Delta_C = y^2(p_1x+p_2)^2 + y\left[x\left(p_2^2-q^2+M_1^2-M_2^2\right)+M_2^2-M_3^2-p_2^2\right] + M_3^2, \quad \text{(A.16b)}$$

$$\text{div} = \frac{2}{\epsilon} - \gamma_E + \ln 4\pi + \ln\mu^2. \quad \text{(A.16c)}$$

Here $\epsilon = 4 - D$ is an infinitesimally small parameter in $D$-dimensional integral, $\mu$ and $\gamma_E$ are arbitrary dimensionful parameter and the Euler constant respectively. div becomes divergent in the limit $D \to 4$, $\epsilon \to 0$. Here we adopt the following shorthand notations for brevity:

$$B_i(p^2; M_A, M_B) = B_i(p^2; A, B)\,,$$
$$C_{ij}(p_1^2, p_2^2, q^2; M_A, M_B, M_C) = C_{ij}(A, B, C)\,.$$

A.5 Form factors

$$F^A_{Z,S} = \frac{N_S}{16\pi^2 v c_W}\big[\lambda_{H^+S^-S_R}[(2-4s_W^2)C_{24}(S_R, S^+, S^+) - 2C_{24}(S^+, S_I, S_R) + s_W^2 B_0(q^2; S^+, S_R)] + \lambda_{H^+S^+S_I}[(2-4s_W^2)C_{24}(S_I, S^+, S^+) - 2C_{24}(S^+, S_R, S_I) + s_W^2 B_0(q^2; S^+, S_I)]\big]\,, \tag{A.18a}$$

$$F^B_{Z,S} = \frac{N_S s_W^2}{16\pi^2 v c_W}\big[\lambda_{H^+S^-S_R}\big(B_0(q^2, S^+, S_R) + 2B_1(q^2, S^+, S_R)\big) + \lambda_{H^+S^-S_I}\big(B_0(q^2, S^+, S_I) + 2B_1(q^2, S^+, S_I)\big)\big]\,, \tag{A.18b}$$

$$G^A_{Z,S} = \frac{N_S M_W^2}{16\pi^2 v c_W}\big[\lambda_{H^+S^-S_R}[(2-4s_W^2)(C_{12}+C_{23}) \times (S_R, S^+, S^+) - 2(C_{12}+C_{23})(S^+, S_I, S_R)] + \lambda_{H^+S^-S_I}[(2-4s_W^2)(C_{12}+C_{23})(S_I, S^+, S^+) - 2(C_{12}+C_{23})(S^+, S_R, S_I)]\big]\,, \tag{A.18c}$$

$$F^A_{\gamma,S} = \frac{N_S s_W}{16\pi^2 v}[\lambda_{H^+S^-S_R}[B_0(q^2, S^+, S_R) - 4C_{24}(S_R, S^+, S^+)] + \lambda_{H^+S^-S_I}[B_0(q^2, S^+, S_I) - 4C_{24}(S_I, S^+, S^+)]]\,, \tag{A.19a}$$

$$F^B_{\gamma,S} = \frac{N_S s_W}{16\pi^2 v}[\lambda_{H^+S^-S_R}\big(B_0(q^2, S^+, S_R) + 2B_1(q^2, S^+, S_R)\big) + \lambda_{H^+S^-S_I}\big(B_0(q^2, S^+, S_I) + 2B_1(q^2, S^+, S_I)\big)]\,, \tag{A.19b}$$

$$G^A_{\gamma,S} = -\frac{4N_S M_W^2 s_W}{16\pi^2 v}\big[\lambda_{H^+S^-S_R}(C_{12}+C_{23})(S_R, S^+, S^+) + \lambda_{H^+S^-S_I}(C_{12}+C_{23})(S_I, S^+, S^+)\big]\,, \tag{A.19c}$$

$$\begin{aligned}
F^A_{Z,\mathrm{2HDM}} = \frac{1}{16\pi^2 v}\bigg[&\lambda_{hH^+H^-}\cos(\beta-\alpha)\bigg\{\frac{s_W^2}{c_W}B_0(q^2, h, H^+) - \frac{2}{c_W}C_{24}(H^+, A, h) + \frac{2c_{2W}}{c_W}C_{24}(h, H^+, H^+)\bigg\}\\
&-\lambda_{HH^+H^-}\sin(\beta-\alpha)\bigg\{\frac{s_W^2}{c_W}B_0(q^2, H, H^+) - \frac{2}{c_W}C_{24}(H^+, A, H) + \frac{2c_{2W}}{c_W}C_{24}(H, H^+, H^+)\bigg\}\\
&+\lambda_{hH^+G^-}\sin(\beta-\alpha)\bigg\{\frac{s_W^2}{c_W}B_0(q^2, h, G^+) - \frac{2}{c_W}C_{24}(G^+, G_0, h) + \frac{2c_{2W}}{c_W}C_{24}(h, G^+, G^+)\bigg\}\\
&+\lambda_{HH^+G^-}\cos(\beta-\alpha)\bigg\{\frac{s_W^2}{c_W}B_0(q^2, H, G^+) - \frac{2}{c_W}C_{24}(G^+, G_0, H) + \frac{2c_{2W}}{c_W}C_{24}(H, G^+, G^+)\bigg\}\\
&+\lambda_{AH^+G^-}\cos(\beta-\alpha)\sin(\beta-\alpha)\\
&\bigg[\lambda_{hH^+H^-}\cos(\beta-\alpha)\bigg\{\frac{s_W^2}{c_W}B_0(q^2, h, H^+)\\
&\frac{2}{c_W}\{C_{24}(G^+, H, A) - C_{24}(G^+, h, A)\}\bigg]\,,
\end{aligned} \tag{A.20a}$$

$$\begin{aligned}
F^B_{Z,\mathrm{2HDM}} = \frac{s_W^2}{16\pi^2 v c_W}
&\times[-\lambda_{hH^+H^-}\cos(\beta-\alpha)\{B_0(q^2, H^+, h) + 2B_1(q^2, H^+, h)\}\\
&+\lambda_{HH^+H^-}\sin(\beta-\alpha)\{B_0(q^2, H^+, H) + 2B_1(q^2, H^+, H)\}\\
&-\lambda_{hH^+G^-}\sin(\beta-\alpha)\{B_0(q^2, G^+, h) + 2B_1(q^2, G^+, h)\}\\
&-\lambda_{HH^+G^-}\cos(\beta-\alpha)\{B_0(q^2, G^+, H) + 2B_1(q^2, G^+, H)\}],
\end{aligned} \tag{A.20b}$$

$$\begin{aligned}
G^A_{Z,\mathrm{2HDM}} = \frac{1}{16\pi^2 v}\,[&\lambda_{hH^+H^-}\cos(\beta-\alpha)M_W^2\\
&\times\bigg\{-\frac{2}{c_W}(C_{12}+C_{23})(H^+, A, h) + \frac{2c_{2W}}{c_W}(C_{12}+C_{23})(h, H^+, H^+)\bigg\}\\
&-\lambda_{HH^+H^-}\sin(\beta-\alpha)M_W^2\\
&\times\bigg\{-\frac{2}{c_W}(C_{12}+C_{23})(H^+, A, H) + \frac{2c_{2W}}{c_W}(C_{12}+C_{23})(H, H^+, H^+)\bigg\}\\
&+\lambda_{hH^+G^-}\sin(\beta-\alpha)M_W^2\\
&\times\bigg\{-\frac{2}{c_W}(C_{12}+C_{23})(G^+, G^0, h) + \frac{2c_{2W}}{c_W}(C_{12}+C_{23})(h, G^+, G^+)\bigg\}\\
&+\lambda_{HH^+G^-}\cos(\beta-\alpha)M_W^2\\
&\times\bigg\{-\frac{2}{c_W}(C_{12}+C_{23})(G^+, G^0, H) + \frac{2c_{2W}}{c_W}(C_{12}+C_{23})(H, G^+, G^+)\bigg\}\\
&+\lambda_{AH^+G^-}\sin(\beta-\alpha)\cos(\beta-\alpha)M_W^2\\
&\times\frac{2}{c_W}\{(C_{12}+C_{23})(G^+, H, A) - (C_{12}+C_{23})(G^+, h, A)\}]\,,
\end{aligned} \tag{A.20c}$$

$$\begin{aligned}
F^{A,F}_{Z,\mathrm{2HDM}} = \frac{2N_t}{16\pi^2 v^2 c_W}
&\times\big[M_t^2\zeta_t\,(v_b + a_b)\,\{4C_{24}(t, b, b) - B_0(q^2, t, b)\\
&- B_0(p_W^2, b, t) - (2M_b^2 - M_Z^2)\,C_0(t, b, b)\}\\
&- M_b^2\zeta_b\,(v_b + a_b)\,\{4C_{24}(t, b, b) - B_0(p_Z^2, b, b)\\
&- B_0(q^2, t, b) - (M_t^2 + M_b^2 - M_W^2)\,C_0(t, b, b)\}\\
&- M_b^2\zeta_b\,(v_b - a_b)\,\{B_0(p_Z^2, b, b) + B_0(p_W^2, t, b)
\end{aligned}$$

$$+ (M_t^2 + M_b^2 - q^2)\, C_0(t,b,b)\}$$
$$+ 2M_t^2 M_b^2 \zeta_t\, (v_b - a_b)\, C_0(t,b,b)\Big]$$
$$+ (M_t, \zeta_t, v_b, a_b) \leftrightarrow (M_b, -\zeta_b, v_t, a_t), \quad \text{(A.21a)}$$

$$F^{B,F}_{Z,\text{2HDM}} = \frac{4 s_W^2 N_t}{16\pi^2 v^2 c_W} \times [M_t^2 \zeta_t\, (B_0 + B_1) - M_b^2 \zeta_b\, B_1](q^2, t, b), \quad \text{(A.21b)}$$

$$G^{A,F}_{Z,\text{2HDM}} = \frac{4 N_c M_W^2}{16\pi^2 v^2 c_W} \times [M_t^2 \zeta_t\, (v_b + a_b)\, (2C_{23} + 2C_{12} + C_{11} + C_0)$$
$$- M_b^2 \zeta_b\, (v_b + a_b)\, (2C_{23} + C_{12})$$
$$- M_b^2 \zeta_b\, (v_b - a_b)\, (C_{12} - C_{11})](t,b,b)$$
$$+ (M_t, \zeta_t, v_b, a_b) \leftrightarrow (M_b, -\zeta_b, v_t, a_t)\,, \quad \text{(A.21c)}$$

$$H^{\text{1PI}}_{Z,F} = \frac{4 N_c M_W^2}{16\pi^2 v^2 c_W} \times \Big[M_t^2 \zeta_t (v_b + a_b)(C_0 + C_{11}) - M_b^2 \zeta_b (v_b + a_b) C_{12}$$
$$+ M_b^2 \zeta_b (v_b - a_b)(C_{12} - C_{11})\Big](t,b,b)$$
$$+ (M_t, \zeta_t, v_b, a_b) \leftrightarrow (M_b, +\zeta_b, v_t, a_t)\,. \quad \text{(A.22)}$$

where,

$$v_f = I_f - s_W^2 Q_f,\; a_f = I_f. \quad \text{(A.23)}$$

$$F^{A}_{\gamma,\text{2HDM}} = \frac{1}{16\pi^2 v} \times [-\lambda_{hH^+H^-} s_W \cos(\beta-\alpha) B_0(q^2, h, H^+)$$
$$+ \lambda_{HH^+H^-} s_W \sin(\beta-\alpha) B_0(q^2, H, H^+)$$
$$- \lambda_{hH^+G^-} s_W \sin(\beta-\alpha) B_0(q^2, h, G^+)$$
$$- \lambda_{HH^+G^-} s_W \cos(\beta-\alpha) B_0(q^2, H, G^+)$$
$$+ 4\, \lambda_{hH^+H^-} s_W \cos(\beta-\alpha) C_{24}(h, H^+, H^+)$$
$$- 4\, \lambda_{HH^+H^-} s_W \sin(\beta-\alpha) C_{24}(H, H^+, H^+)$$
$$+ 4\, \lambda_{hH^+G^-} s_W \sin(\beta-\alpha) C_{24}(h, G^+, G^+)$$
$$+ 4\, \lambda_{HH^+G^-} s_W \cos(\beta-\alpha) C_{24}(H, G^+, G^+)], \quad \text{(A.24a)}$$

$$F^{B}_{\gamma,\text{2HDM}} = \frac{1}{16\pi^2 v} [-\lambda_{hH^+H^-} \cos(\beta-\alpha)$$
$$\times s_W (2\, B_1(q^2, H^+, h) + B_0(q^2, H^+, h))$$
$$+ \lambda_{HH^+H^-} \sin(\beta-\alpha)$$
$$\times s_W (2\, B_1(q^2, H^+, H) + B_0(q^2, H^+, H))$$
$$- \lambda_{hH^+G^-} \sin(\beta-\alpha)$$
$$\times s_W (2\, B_1(q^2, G^+, h) + B_0(q^2, G^+, h))$$
$$- \lambda_{HH^+G^-} \cos(\beta-\alpha)$$
$$\times s_W (2\, B_1(q^2, G^+, H) + B_0(q^2, G^+, H))] \quad \text{(A.24b)}$$

$$G^{A}_{\gamma,\text{2HDM}} = \frac{M_W^2}{16\pi^2 v} [4\, \lambda_{hH^+H^-} s_W$$
$$\times \cos(\beta-\alpha)(C_{12} + C_{23})(h, H^+, H^+)$$
$$- 4\, \lambda_{HH^+H^-} s_W \sin(\beta-\alpha)(C_{12} + C_{23})(H, H^+, H^+)$$
$$+ 4\, \lambda_{hH^+G^-} s_W \sin(\beta-\alpha)(C_{12} + C_{23})(h, G^+, G^+)$$
$$+ 4\, \lambda_{HH^+G^-} s_W \cos(\beta-\alpha)(C_{12} + C_{23})(H, G^+, G^+)] \quad \text{(A.24c)}$$

$$G^{A,F}_{\gamma,\text{2HDM}} = \frac{4 N_c Q_b M_W^2}{16\pi^2 v^2 c_W} \times \Big[M_t^2 \xi_t (2C_{23} + 2C_{12} + C_{11} + C_0)$$
$$- M_b^2 \zeta_b (2C_{23} + C_{12})\Big](t,b,b)$$
$$+ (M_t, \xi_t, Q_b) \leftrightarrow (M_b, -\xi_b, Q_t)\,. \quad \text{(A.25)}$$

$$H^{\text{1PI}}_{\gamma,F} = \frac{4 N_c Q_b M_W^2}{16\pi^2 v^2 c_W} \times [M_t^2 \zeta_t (C_0 + C_{11}) - M_b^2 \zeta_b C_{12}$$
$$+ M_b^2 \zeta_b (C_{12} - C_{11})](t,b,b)$$
$$+ (M_t, \zeta_t, Q_b) \leftrightarrow (M_b, +\zeta_b, Q_t) \quad \text{(A.26)}$$

$$F_Z = F^A_{Z,S} + F^B_{Z,S} + F^A_{Z,\text{2HDM}} + F^B_{Z,\text{2HDM}} + F^{A,F}_{Z,\text{2HDM}}\,, \quad \text{(A.27a)}$$
$$G_Z = G^A_{Z,S} + G^A_{Z,\text{2HDM}} + G^{A,F}_{Z,\text{2HDM}}\,, \quad \text{(A.27b)}$$
$$H_Z = H^{\text{1PI}}_{Z,F}\,, \quad \text{(A.27c)}$$
$$F_\gamma = F^A_{\gamma,S} + F^B_{\gamma,S} + F^A_{\gamma,\text{2HDM}} + F^B_{\gamma,\text{2HDM}} + F^{A,F}_{\gamma,\text{2HDM}}\,, \quad \text{(A.27d)}$$
$$G_\gamma = G^A_{\gamma,S} + G^A_{\gamma,\text{2HDM}} + G^{A,F}_{\gamma,\text{2HDM}}\,, \quad \text{(A.27e)}$$
$$H_\gamma = H^{\text{1PI}}_{\gamma,F}\,. \quad \text{(A.27f)}$$